\documentclass[a4paper,12pt]{report}
\usepackage{graphicx}
\usepackage{titling}
\usepackage{amsmath}
\usepackage[export]{adjustbox}
\usepackage{subcaption}
\usepackage[english]{babel}
\usepackage[numbers]{natbib}
\usepackage{dirtytalk}
\usepackage{hyperref}
\usepackage{float}
\usepackage{pdfpages}
\usepackage{amssymb}
\usepackage{titlesec}
\usepackage{algorithm}
\usepackage[noend]{algpseudocode}
\usepackage[utf8]{inputenc}
\usepackage{multirow}
\usepackage{chngcntr}
\usepackage{textcomp}
\counterwithout{footnote}{chapter}
\usepackage[format=plain,
            labelfont={bf,it},
            textfont=it]{caption}  
  
\def\blankpage{%
      \clearpage%
      \thispagestyle{empty}%
      \addtocounter{page}{-1}%
      \null%
      \clearpage}

\titleformat{\chapter}[display]
  {\normalfont\bfseries}{}{0pt}{\Huge}
  
\usepackage[margin=2cm]{geometry}
\newcounter{equationset}

\begin{document}

\begin{titlepage}
    \begin{center}
        \centering
    	\noindent\makebox[\linewidth]{\rule{1.5\paperwidth}{0.4pt}}
    	\smallskip
    	
    	\includegraphics[width=0.5\textwidth]{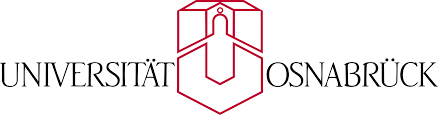} \par \vspace{0.5cm}\par
    	{\scshape Bachelor Thesis\par}
    	\noindent\makebox[\linewidth]{\rule{1.5\paperwidth}{0.4pt}}
        
        \vspace{2cm}    
        \Huge
        \textbf{Teaching a Machine to Diagnose a Heart Disease}
            
        \vspace{0.5cm}
        \LARGE
        Beginning from digitizing scanned ECGs \\
        to detecting the Brugada Syndrome (BrS)
        \vspace{1cm}
        
        \includegraphics[width=0.55\textwidth]{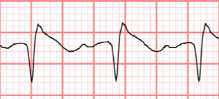}
        
        \vspace{3cm}
            
        \textbf{Simon Jaxy}
            
        \vfill
        
        \Large
        Cognitive Science\\
        University of Osnabrück\\
        Germany\\
        \today
            
    \end{center}
\end{titlepage}

\blankpage

\newpage
\pagenumbering{gobble}
\begin{center}
\Large{\textbf{Teaching a Machine to Diagnose a Heart Disease}}
	\par \vspace{0.3cm}
	\textit{Beginning from digitizing scanned ECGs} \\
	\textit{to detecting the Brugada Syndrome (BrS)} \par
	\vspace{2.5cm}
\large{Simon Jaxy}\par
{sjaxy@uos.de \par
$969281$ \par
Bachelor's programme in Cognitive Science \par
\vspace{2.5cm}
Supervised by:
\par
\textit{Nico Potyka (University of Osnabrück) \\ 
        Isel Del Carmen Grau Garcia
        (Vrije Universiteit Brussel)}
\par
\vspace{2.5cm}
In collaboration with: \par
\textit{VUB Artificial Intelligence Lab}}
\par
\vfill
{\large \today\par}
\end{center}

\newpage
\pagenumbering{roman}

\chapter*{Declaration of Authorship}

\noindent I, Simon Jaxy, declare that this thesis titled, "Teaching a Machine to Diagnose a Heart Disease. Beginning from digitizing scanned ECGs to detecting the Brugada Syndrome (BrS)" and the work presented in it are my own. I confirm that:

\begin{itemize} 
\item This work was done wholly or mainly while in candidature for a research degree at this University.
\item Where any part of this thesis has previously been submitted for a degree or any other qualification at this University or any other institution, this has been clearly stated.
\item Where I have consulted the published work of others, this is always clearly attributed.
\item Where I have quoted from the work of others, the source is always given. With the exception of such quotations, this thesis is entirely my own work.
\item I have acknowledged all main sources of help.
\item Where the thesis is based on work done by myself jointly with others, I have made clear exactly what was done by others and what I have contributed myself.\\
\end{itemize}
 
\noindent Signed:\\
\rule[0.5em]{25em}{0.5pt} 
 
\noindent Date:\\
\rule[0.5em]{25em}{0.5pt} 

\newpage

\begin{abstract}
    Medical diagnoses can shape and change the life of a person drastically. Therefore, it is always best advised to collect as much evidence as possible to be certain about the diagnosis. Unfortunately, in the case of the Brugada Syndrome (BrS), a rare and inherited heart disease, only one diagnostic criterion exists, namely, a typical pattern in the Electrocardiogram (ECG). \newline
    
    In the following treatise, we question whether the investigation of ECG strips by the means of machine learning methods improves the detection of BrS positive cases and hence, the diagnostic process. We propose a pipeline that reads in scanned images of ECGs, and transforms the encaptured signals to digital time-voltage data after several processing steps. Then,  we present a long short-term memory (LSTM) classifier that is built based on the previously extracted data and that makes the diagnosis. 
    \newline
    
    The proposed pipeline distinguishes between three major types of ECG images and recreates each recorded lead signal. Features and quality are retained during the digitization of the data, albeit some encountered issues are not fully removed (Part I). Nevertheless, the results of the aforesaid program are suitable for further investigation of the ECG by a computational method such as the proposed classifier which proves the concept and could be the architectural basis for future research (Part II). 
    \newline
    This thesis is divided into two parts as they are part of the same process but conceptually different. 
    \newline 
    
    It is hoped that this work builds a new foundation for computational investigations in the case of the BrS and its diagnosis.
\end{abstract}

\newpage
\setcounter{page}{3}
\tableofcontents
\listoffigures 
\listoftables 
\listofalgorithms

\newpage
\pagenumbering{arabic}
\chapter{Preface}
The dissertation Teaching a Machine to Diagnose a Heart Disease" that now lies in front of you, presents a guideline of how a diagnosis over a severe disease can be made starting only from scanned images of ECG recordings. It has been created to fulfill the graduation requirements of the Cognitive Science BSc. program at the University of Osnabrück. The writing process lasted from December 2019 until the beginning of March 2020. 
\newline

The here proposed work is part of the multi-disciplinary IMAGica project (VUB IRP8) that seeks to provide an improved and effective diagnostic process for patients. In particular, the research conducted was undertaken by the request of the VUB Artificial Intelligence Lab together with the UZ Brussel hospital that both play a major role in the IMAGica project. The studies related to this project are approved by the Medical Ethics Committee UZ Brussels/VUB. My research question was formulated together with my supervisor Isel del Carmen Grau Garcia and with this treatise it is hoped to lay the first foundation of a deeper investigation of the Brugaga Syndrome. 
\newline

I am thankful to both of my supervisors, Isel del Carmen Grau Garcia and Nico Potyka, for allowing me to become part of this project and letting me write my thesis about it as well as giving me excellent guidance throughout the process.  
\newline

Furthermore, I want to thank my colleagues in the VUB AI Lab for the discussions and the resulting ideas that emerged from this process. I acknowledge my family and friends for giving me strength and support to endure the creation process and  I am especially grateful for the support of my companion as she was a skilled "sparring partner" when discussing concepts and ideas. 
\newline

Simon Jaxy 
\newline

Brussels, \today

\newpage
\chapter{Part I: The Digitization Process}
\section{Introduction}
In this part of the thesis, we propose an automated pipeline that faces the challenge of digitizing ECG signals. Hereby, the workflow of the pipeline is presented in detail as well as a schematic description. During the process, it is most effective to separate different types of ECG images into different streams. This distinction is thematized later in the discussion part.
\newline

First, we state the problem and our objective. Related work is presented subsequently. Then, we propose the architecture of the pipeline in chapter three and evaluate it in chapter four. Finally, as already hinted, a discussion part is added.	

\subsection{Problem Statement}
The ECG is a physiological method to inspect the heart. It does so by converting the electrical activity of the human heart into a graphical representation, usually printed onto a reference grid. Often, it is one of the very first methods applied to investigate the cardiovascular system as different impairments, such as structural heart diseases, ischemic heart diseases or other causes of symptoms that lie outside the heart, can be detected by the ECG \cite{Woudstra2015}.\newline 

During the ECG recording, the electrical activity is measured against time \cite{Wasilewski2012}. Electrodes are placed onto the chest and limb which record the change of the electrical potential throughout depolarization and repolarization of the heart. "The sources of the electrical potentials are contractile cardiac muscle cells \cite[p.1]{Wasilewski2012}", hence, the heartbeat causes the electrical activity. In this non-invasive technique, the order and placement of the electrodes are of utmost importance for the accuracy of the recorded signals \cite{Woudstra2015}.\newline 

For many hospitals, digitized ECG signals are not the standard as they rely on the printed version, due to the high costs of modernizing the equipment. Large storage spaces are needed to store patient data, especially in very high populated regions, e.g., India or China  \cite{5478930}. Fast access to the medical records of patients cannot be guaranteed and hence, a call for fast accessible data is evolving \cite{5478930},\cite{Garg2012}. Moreover, the ECG is printed onto a specific thermal paper, which is not storable without harming the quality of the signal \cite{Kleaf2015}.    \newline 

The need for the digitization of the ECG lies at hand. For one, the longtime storage of the recorded ECGs would not result in losing information. Plus, a fast transfer of the data would be possible when passing information from one medical institute to another. Furthermore, modern computational methods can be applied to investigate the recorded signals. Digitizing the ECG will also lead to a vast increase in the quantity of data that is available for research purposes as old ECGs can be recovered and newer ones additionally recorded. This will become especially relevant when dealing with rare heart conditions for which not many known cases exist but the records could be gathered over many years in the past.

\subsection{Objective}
Digitizing the ECG is a many step process that places a few challenges for the human engineer. Also, the quality of the scan plays an important role because the scanned image might be tilted, etc. Therefore, it is mandatory to design a pipeline that takes these challenges into account. \newline

This pipeline aims to convert the pixel information obtained from the scanned images into a one-dimensional vector that expresses the ECG signal such that it can be easily reconstructed from the vector. Moreover, the goal is to reconstruct every lead that is visible on the sheet. Not taken into account is personal information from the patient, which could also be accessed from the paper. The reconstruction is implemented via Python \cite{Python3} 3.7.3 and with the help of the libraries OpenCV \cite{opencv_library}, Numpy \cite{Numpy}, Scipy \cite{2020SciPy-NMeth}, Pandas \cite{Pandas} and Matplotlib \cite{matplotlib}.

\section{Related Work}
In the following, we present different solutions to the digitization problem as they are presented in the literature as well as their validation methods. 

\subsection{Pipelines}
A variety of different solutions were introduced over the last decades, proposing different ways of digitizing ECG signals (the entire stream of the process is called pipeline). Most of them focus on a human-engineered solution, e.g., see \cite{5478930},\cite{Garg2012},\cite{Kleaf2015}. Waits and Soliman \cite{Waits2017} present an overview of different human-engineered techniques\footnote{The technique is human engineered in a sense that on every step and every parameter is chosen by a human as opposed to, e.g., a deep learning method.}. The aim is always to design a program that is as automated as possible by reducing the amount of manually interfering in the process. \newline
Generally, the workflow goes as described by Waits and Soliman \cite{Waits2017}. Firstly, the recorded ECG is scanned and possible rotations and skewing methods are applied.  Then, the background grid is removed. Afterward, different kinds of processing techniques are applied to refine the signal. The final step that follows is the extraction of the signal. From time to time an optical character recognition is additionally performed to obtain valuable patient information. \newline	
Recently, Fontanarava \cite{Fontanarava2019} introduced a deep learning method for fully automated ECG feature extraction. It proposes the usage of three different Convolutional Neural Networks, responsible for layout detection, column-wise signal segmentation, and signal retrieval. This solution is more liberal than compared to a human-engineered one as it allows for quicker adaption between different kinds of ECG images and once the architecture is fully built, the amount of human interference is reduced to a minimum.
\newline
For our purposes, we construct a human-engineered solution that orients itself on the works of Waits and Soliman \cite{Waits2017}.

\subsection{Validation}
As described by Waits and Soliman \cite{Waits2017} and further summarized by Fontanarava \cite{Fontanarava2019}, no standard technique or metric has been established yet to validate the quality of the recovered ECG signal. There are almost as many comparison methods introduced as there are different types of humanly engineered solutions to the task. A fit between the original and recovered signal combined with a correlation between the signals was suggested by Ravichandran et al. \cite{Ravichandran2013}, whereas Badilini et al. preferred to use a least-square fit analysis \cite{Baldini2005}. While others, e.g., Wand and Mital \cite{Wang1996}, rely on visual comparison only. 
\newline

At the time of writing this thesis, we did not have any already digitalized ECGs from the Brugada Syndrome, the study case at hand. Therefore, we could not make use of metrics to verify the quality of our extracted signal.
Out of necessity, we decided to rely on graphical comparisons as our validation method, which is described in the fourth chapter. 

\section{Pipeline}
The available records of ECGs for detecting BrS are limited and diverse in age, source and quality of the image. A uniform process for different kinds of ECGs manifested itself to be not efficient as the obtained images differed too much to undergo the same workflow. Therefore, a distinction between three types of images is made. The first type (Type 1) is composed of newer ECGs of higher quality with a colored background grid. The second type (Type 2) consists of older ECG images that were written in a manner such that the grid and the ink of the signal have the same color and intensity leading to an almost binary image. Lastly, the third type (Type 3) are images which are also binary but without a background grid. Distinguishing between the types results in a separated flow that is intersected at several procedural steps where only parameters deviate. \newline 

The general procedure goes as follows: First, the images are pre-processed to rotate (if necessary) and downsize the image. Then, an intermediate step of obtaining the inner frame image from Type 2 and Type 3 ECGs follows. Next, thresholding is executed to binarize the image and a grid removal is performed. Images are further cleansed from noise and other perturbations. Finally, the signal is extracted, mapped into time-voltage coordinates and effectively upsampled.

\begin{figure}[H]
    \begin{subfigure}{\linewidth}
    \includegraphics[scale=0.50]{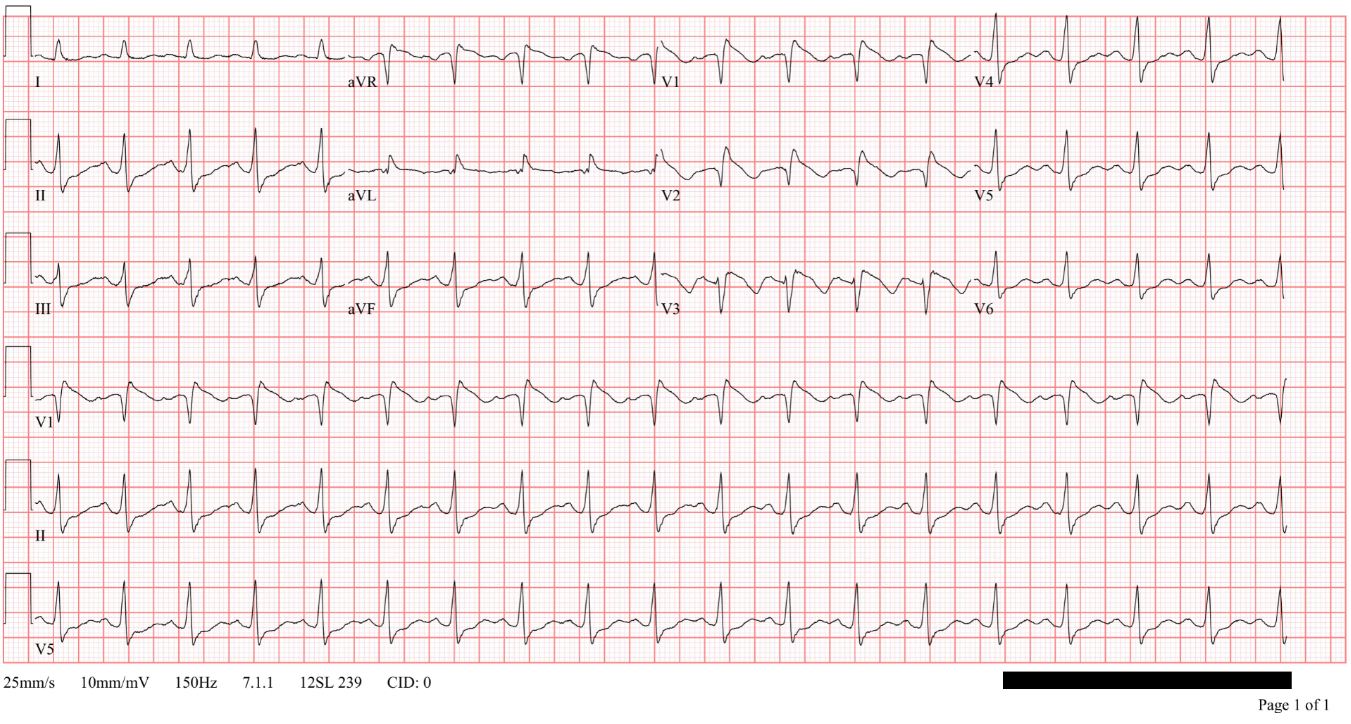}
    \centering
    \caption{}
    \end{subfigure}\par\medskip
    \begin{subfigure}{\linewidth}
    \includegraphics[scale=0.50]{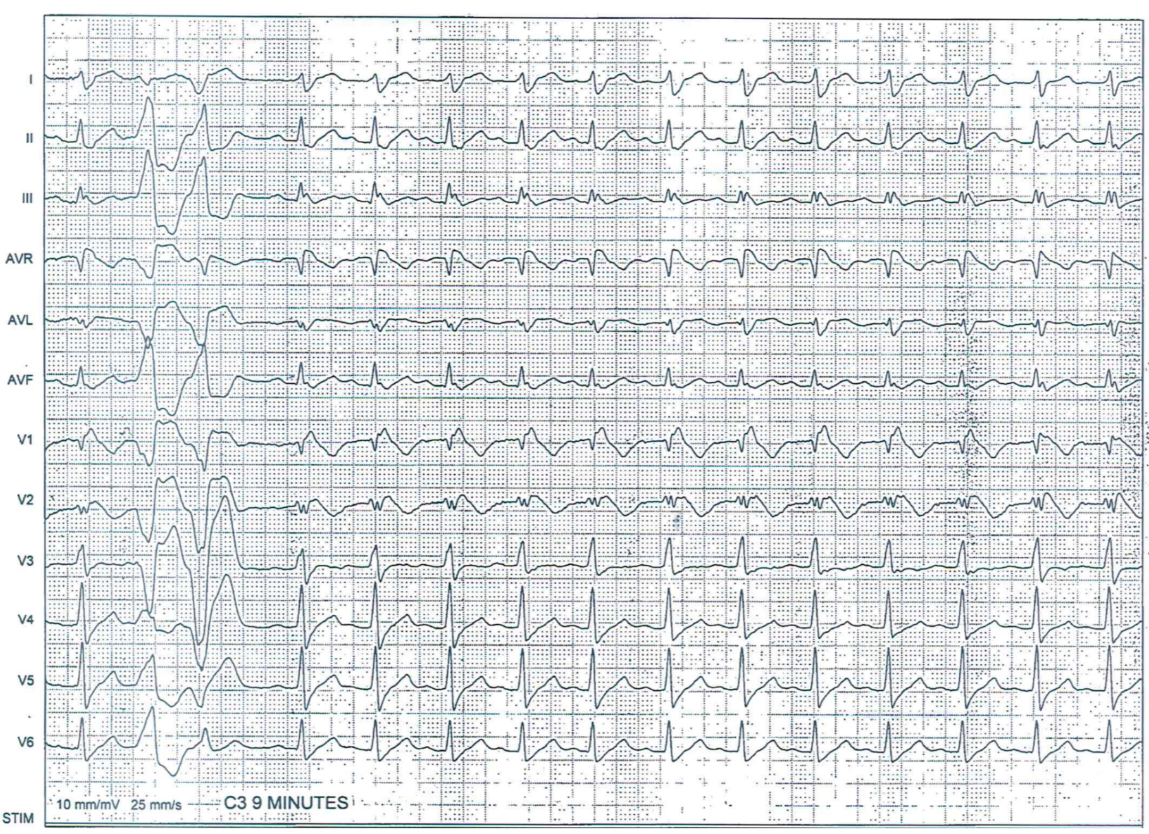}
    \centering
    \caption{}
    \end{subfigure}\par\medskip
    \begin{subfigure}{\linewidth}
    \includegraphics[scale=0.50]{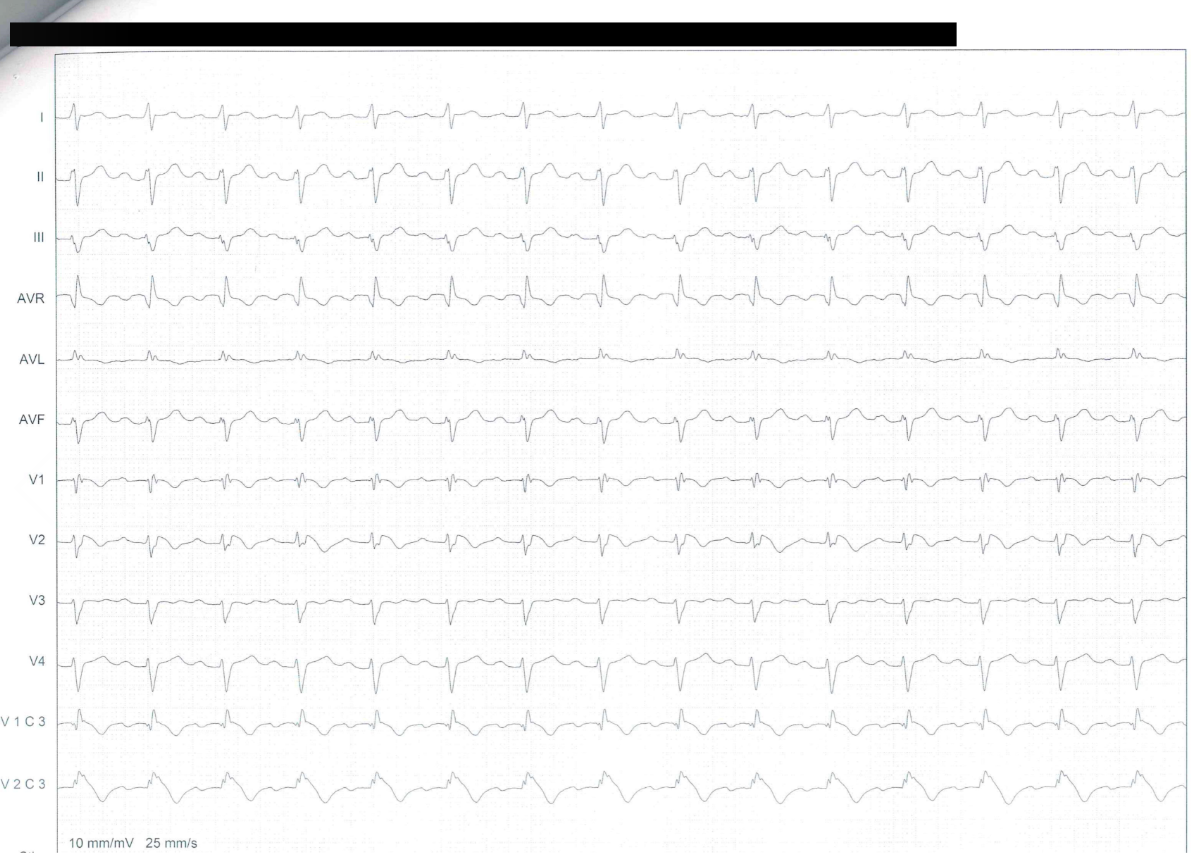}
    \centering
    \caption{}
    \end{subfigure}
    \caption{The three different types of ECG sheets (a) Type 1, (b) Type 2, (c) Type 3.}
    \label{fig:mesh1}
    \centering
    
\end{figure}

\newpage
\begin{figure}[H]
    \includegraphics[max size={\textwidth}{\textheight}]{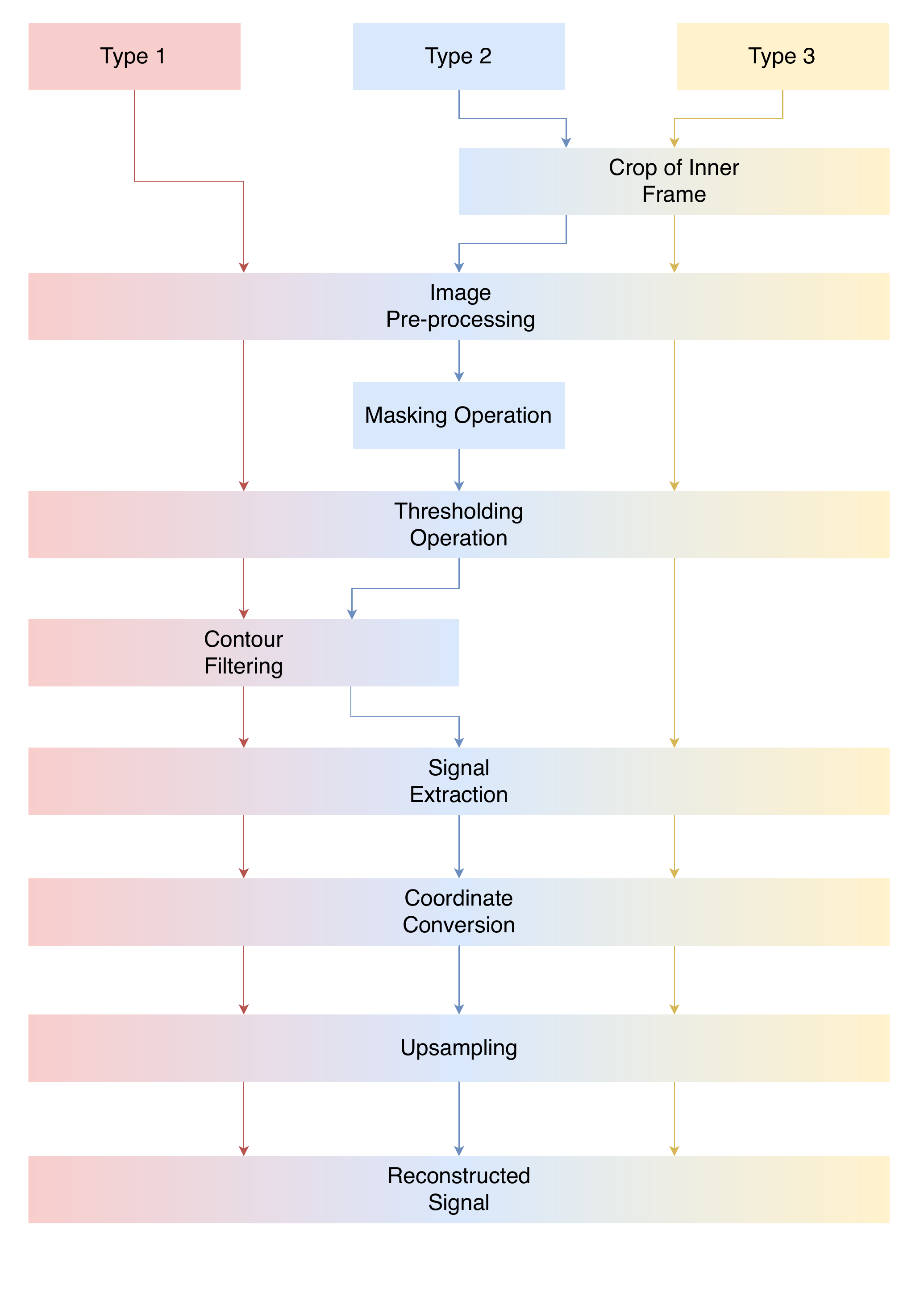}
    \caption{Digitization Pipeline}
    \centering
\end{figure}
\newpage

\subsection{Pre-processing}
Before the processing starts, the image has to be converted into a grayscale mode. Previously colored pixels are hence mapped to intensity pixels representing the amount of light of each color \cite{Grayscale2019}. \newline
The first step of the pipeline is to rotate the images in case they have not been placed correctly into the scanner. This is important since the retrieved signal should not be tilted but rather represent the originally recorded signal. The idea of the process is that detected grid-edges in the image can be combined to lines. By calculating the slopes of the found lines, the necessary rotation angle can be computed. 
Automatic \cite{Rosebrock2015} Canny edge detection \cite{Canny:1983:FEL:889502} algorithm provides a way to detect the edges in the image. First, a noise reduction is performed to obtain a cleaner image by using a low-pass filter (usually a Gaussian Kernel) \cite{Shrivakshan2012}. Then, the edge gradients are calculated by computing the first derivative in the horizontal and vertical direction since \say{edges occur at locations of \textit{steepest slopes} \cite[p.239]{Szeliski2010}}. 

\begin{figure}[H]
\begin{align}
& J_\sigma=\nabla[G_\sigma(\mathbf{x})\ast I(\mathbf{x})]=[\nabla G_\sigma](\mathbf{x})\ast I(\mathbf{x}) \label{eq:001}\\
& \nabla[G_\sigma](\mathbf{x}) = (\frac{\partial G_\sigma}{\partial x}, \frac{\partial G_\sigma}{\partial y})(x) = [-x -y]\frac{1}{\sigma^3}exp(-\frac{x^2+y^2}{2\sigma^2}) \label{eq:002}
\end{align}
\end{figure}

The smoothing operation and subsequent gradient calculation \cite[p.240]{Szeliski2010} are depicted in \ref{eq:001} and \ref{eq:002}, where $J_\sigma$ is the gradient of the smoothed image, $G_\sigma$ the Gaussian kernel function and $I$ the original image. $\mathbf{x}$ is the input , $x$ denotes the horizontal coordinate, $y$ stands for the vertical coordinate and $\sigma$ is the width of the Gaussian. 
\newline 
From the equation follows that the input image is convolved \say{with the horizontal and vertical derivatives of the Gaussian kernel function \cite[p.240]{Szeliski2010}} to smooth the image and find its edge gradients.
\newline

Next, every pixel in the image is tested, whether or not it is a local maximum in its neighborhood by \say{locating the zero-crossings in the second derivative in the gradient direction \cite[p.50]{Canny:1983:FEL:889502}}. Lastly, a double thresholding operation (including a low threshold and a high threshold) decides which pixels are edges and which are not. A pixel above the high threshold is taken as immediate output, \say{as is the entire connected segment of the contour which contains the point and which lies above a low threshold \cite[p.60]{Canny:1983:FEL:889502}}, while the pixels with an intensity gradient below the low threshold are discarded as none-edges. For our purposes, the zero parameter implementation that was established by Rosebrock \cite{Rosebrock2015} shows itself to be reasonable because it facilitated the search for a suited parameter. \newline

Using the detected edges, the probabilistic Hough transform (fully described in the OpenCV documentation \cite{cvHough}), is applied in order to find connected lines within the images by performing a search for the describing parameters. First each point $(x,y)$ of the edge that is described by $ y= mx + c$ (and was previously detected by the edge detection algorithm) can be expressed in polar coordinates $(\rho, \theta)$, where $\rho$, computed via the equation $\rho = x\cos{\theta}+y\sin{\theta}$, is the distance from the origin and $\theta$ the angle of the line and the horizontal axis. A line is found in coordinate-space if the line $\rho=x_i\cos{\theta}+y_i\sin{\theta}$ of a coordinate-pair $(x_i,y_i)$ crosses another line $p=x_j\cos{\theta}+y_j\sin{\theta}$ of the coordinate-pair $(x_j, y_j)$ in the parameter space $(\rho, \theta)$. An array of $\rho$ (rows) and $\theta$ (columns) (aka accumulator) is initiated. Each $\rho$ is computed by plugging in the points $(x_i,y_i)$ into the line equation of the parameter space together with a $\theta$-value that is iterated over $\theta \in (1\dots180)$. For each $(\rho,\theta)$-pair, a one is added to the value in the corresponding array cell. This process is repeated for every point on the line of the coordinate-space. Now, the $(\rho,\theta)$-cell with the maximum value gives indication where the line lies with parameters $(\rho,\theta)$.
\newline
\smallskip

\textit{The summary of the pre-processing steps:} 
\begin{algorithm}[H]
\caption{Pre-process flow}
\begin{algorithmic}[H]
\Require{ImageList}
\For {$image \in imageList$}
\State $edges \gets Canny(image)$
\State $lines \gets ProbabilisticHoughTransform(edges)$
\State $angleList \gets []$
    \For {$i, line \in enumerate(lines)$}
    \State $angle \gets (y_1-y_2) / (x_1-x_2) * $180$/\pi$
    \State $angleList[i] \gets angle$
    \EndFor
\State $medianAngle \gets median(angleList)$
\State $image \gets Rotate(image, medianAngle)$ 
\State $image \gets Downsize(image)$
\EndFor
\end{algorithmic}
\label{alg:01}
\end{algorithm}
\smallskip

The probabilistic version differs in the sense that it only considers a random subset of points that are enough for performing the line detection and thus yielding a more computationally-efficient algorithm \cite{cvHough}.  
\newline
By performing the algorithm, starting $(x_1, y_1)$ and endpoints $(x_2, y_2)$ of a line are returned as the output. An angle for rotation can now be determined by calculating the angle of the slope of the line $angle = (y_1-y_2)/(x_1-x_2) * 180/\pi$ (the angle is thus transformed from radiance into degrees). Depending on whether a selected line is horizontal or vertical a shift is added. All angles are saved and then the median angle is chosen.
\newline
The entire scheme was tested using 851 randomly rotated images and inspecting the rotation outcome with the eye. Of the 851 images, 800 were rotated correctly, yielding an accuracy of 94\%.  
\newline

Lastly, the images needs to be downsized to reduce the computational effort. Type 1 and Type 2 ECGs are shrunk to a quarter of their original size and $(594$x$1042)$, whereas Type 3 ECGs are reduced to an eighth of their previous size\footnote{To illustrate, three exemplary ECG images from Type 1 to Type 3 are downsized from $(2834$x$5313)$, $(2375$x$4167)$, $(3910$x$5875)$ to  $(708$x$1328)$, $(594$x$1042)$ and $(489$x$734)$ respectively.}. During this process, pixels are resampled by taking the pixel to area relationship \cite{GEOCV} into account such that a weighted average for the neighborhood in the defined area is calculated and taken as the new output pixel for this area \cite{MCHUGH20}.

\subsection{Inner Frame}
The signals of Type 2 and Type 3 ECGs are located inside a black frame, see Figure \ref{fig:mesh1}. Cropping the inner picture provides a good means to retain the quality of the signal as the black frame is interfering with some of the signals during the processing step. Therefore, the first step is to yield the inner image. Type 1 ECGs do not need to be processed this way since their frame is in a different color which makes it possible to erase it simply by a thresholding operation. Thresholding means that a constant (threshold) is determined that binarizes the image such that all pixels whose value lies under the threshold will be set to be black and every pixel above the threshold will be set to be white. 	
\newline 
\smallskip

\textit{The entire process of extracting the inner frame:}
\begin{algorithm}[H]
\caption{Extracting inner frame. Note: ImageList only contains images from Type 2 and Type 3 ECGs. $\ast$ denotes the convolution operation.}
\begin{algorithmic}[H]
\Require{ImageList}
\For {$image \in imageList$}
\State $copyImage \gets copy(image)$
\State $blurredImage \gets medianFilter(copyImage, 5)$
\State $sharpenedImage \gets blurredImage \ast sharpeningKernel$
\State $binaryImage \gets Treshold(sharpenedImage)$
\State $closedImage \gets MorphologicalClosing(binaryImage$
\State $contourList \gets FindContours(closedImage)$
\State $maxContour \gets {max \atop area}(contourList)$
\State $(y,x,h,w) \gets Coordinates(maxContour)$
\State $(y,x,h,w) \gets (x,y,h,w) + coordinateShift$
\State $image \gets image[y:h,x:w]$
\EndFor
\end{algorithmic}
\label{alg:02}
\end{algorithm}
\smallskip

A copy of the original image is made to secure that the image stays unharmed during this process. All of the following image manipulation techniques are applied to the copy to ensure that the signals are kept as they are. \newline
Finding the frame algorithmically is achieved by finding the contours inside the image. Contours can be seen as curves that unite connected points having the same color or intensity \cite{Contours2013}. 
In pursuance of the contour seeking, some pre-processing steps have to be applied to the copy. First of all, the image is blurred to reduce high-frequency noise. This is achieved by using a median-filter, \say{which selects the median value from each pixel's neighborhood \cite[p.124]{Szeliski2010}}, in this case, a 5x5 neighborhood. Afterward, the image is sharpened again to strengthen the inked pixels by convolving the blurred image with a sharpening kernel\footnote{the kernel is of the form: $\begin{bmatrix}-1 & -1 & -1 \\ -1 & 9 & -1 \\ -1 & -1 & -1 \end{bmatrix}$ which was found empirically after being inspired by \cite{ImageKernel}.}. Then, the copy is binarized using a threshold operation. A morphological closing operation ensures the continuity of connected pixels\footnote{Morphological operations are nonlinear operations that are performed on binary images. To execute the operation, \say{we first convolve the binary image with a binary \textit{structuring} kernel and then select a binary output value depending on the threshold result of the convolution \cite[p.126]{Szeliski2010}. The convolution is described by $c=f\otimes s$, where $c$ represents the integer-valued count of number 1s (foreground-pixels) inside the binary kernel as it is convolved with the image, $f$ stands for a binary image and $s$ resembles the kernel. Morphological close is then defined as $close(f,s)=erode(dilate(f,s),s)$, with $dilate(f,s)=\theta (c,1)$, $erode(f,s)= \theta (c,S)$}, where $S$ is the size of the structuring element and $\theta = \begin{cases} 1 & \text{if $f \geq t$} \\ 0 & \text{else} \end{cases}$, where $t$ stands for the threshold-value, as it is further described in \cite[p.128f]{Szeliski2010}.}. Finally, the contours can be found. The algorithm that we use had been first presented by Suzuki and Abe \cite{SUZUKI198532}. It finds borders within the binarized image and follows them along to determine the structure of connected pixels. The connectivity of pixels are judged by either taking into account the 4- or 8-neighborhood of the pixel. A pixel at point $(i,j)$ is said to be a boarder pixel if there exists a 0-valued (background) pixel in its respective neighborhood. The image is scanned and at each border pixel, the point is labeled. Its border is followed until the entire border is found. If two labels meet, one of them consumes the other together with its entire border. After following an entire border, the image scan is continued where it had been interrupted. A hierarchical structure is thus built by labeling found boarders from outer to inner borders that are separated by holes (areas in which no labeled pixel exists). Using the contour-information, the rectangle with the biggest pixel area is found and its respective coordinates. These coordinates are then used to crop the original (not processed copy-image) such that the result is the inner-frame-image. Albeit, sometimes, the found rectangular is not precise. 
\newline
In pursuance of a generalized pipeline that works for many images, a coordinate shift has to be applied to ensure that no grid line was visible anymore. Unfortunately, for some ECG this leads to a loss of information as some of the signals is cut out. Additionally, Type 2 and Type 3 images have different shifts as they diverged too strongly in their frame placement.  For two cases of the Type 3 images, the shifting-coordinates have to be found manually after applying the procedure described above since the algorithm did not find the framing borders. 

\subsection{Thresholding}
Following these processing steps, a thresholding operation is performed on all types of ECG images to detach signals from the rest of the image. The thresholding operation $\theta = \begin{cases} 1 & \text{if $p \geq t$} \\ 0 & \text{else} \end{cases}$, where $p$ is a pixel-intensity-value and $t$ the threshold-value, results in binarized images. $t$ differs depending on which type of ECG image was operated on. Binarizing the images ensures that the signals are set as foreground and the rest, e.g. grid lines, is set as background and thus erased. 

\subsection{Contour Filtering}
However, the thresholding operation does not result in isolated signals in the case of Type 1 and Type 2 ECGs. In the case of the Type 1 images, characters, describing which channel is being depicted, remain that need to be removed while Type 2 images are contaminated by leftovers of the background grid that could not be fully erased in the thresholding and convolution procedure. \newline
To localize and remove unwanted groups of pixels we apply the aforementioned finding contour procedure. This time, an integer value is determined empirically which acts as a threshold representing the minimum amount of pixels that need to be enveloped by the detected contour. Contours of characters or leftover grids are filtered using this process albeit the Type 2 ECGs are harmed during this process because the grid residuals have not been entirely separable from the signal. A trade-off appears where the decision has to be made whether to lose parts of the signal or falsify the quality of the signal by including parts of the grid. For the sake of having the best quality while still preserving most signals, a compromise is achieved in which minor parts of the signals are lost while only minimally tampering the signal. 	

\subsection{Signal Extraction}
As the goal is to recreate each signal, a way has to be found to establish a full separation of the signals such that the information about each channel can be stored and accessed independently. One of the major problems during this procedure is to isolate overlapping signals which occurred quite frequently due to high peaks within the ECG signal. It is mandatory to reduce the loss in quality by the separation process. Therefore, a height-coordinate has to be discovered that leads to the smallest disruption of the signal. \newline

\begin{figure}[H]
    \hspace*{-0.5cm}
    \centering
    \begin{minipage}{0.48\textwidth}
        \centering
        \includegraphics[width=0.9\textwidth]{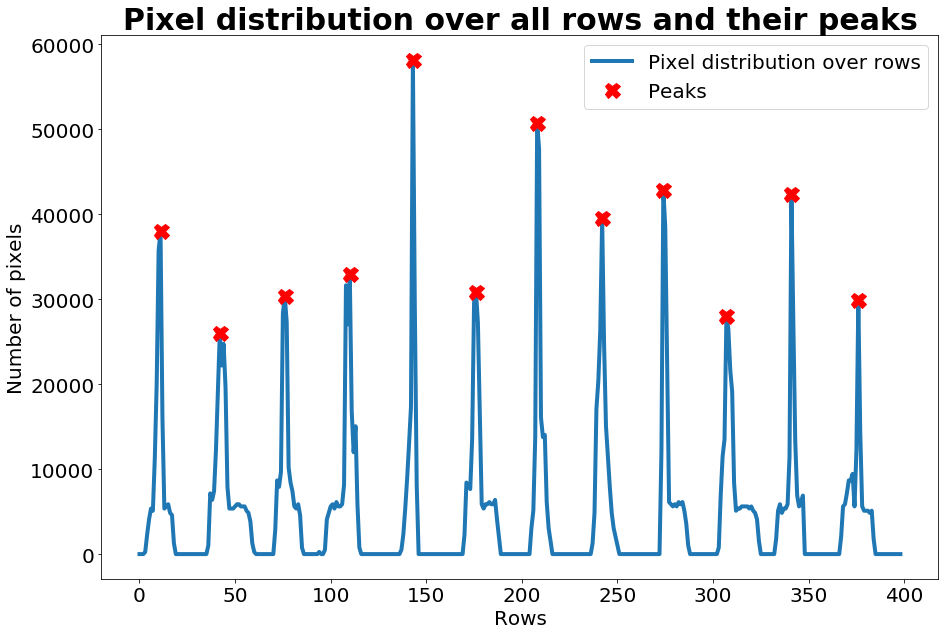} 
        \caption{For every row the amount of pixels in it is plotted together with peaks indicated where the maxima lie. (Example of a Type 3 ECG).}
        \label{fig:mesh04}
    \end{minipage}
    \hspace*{+0.5cm}
    \begin{minipage}{0.48\textwidth}
        \centering
        \includegraphics[width=0.9\textwidth]{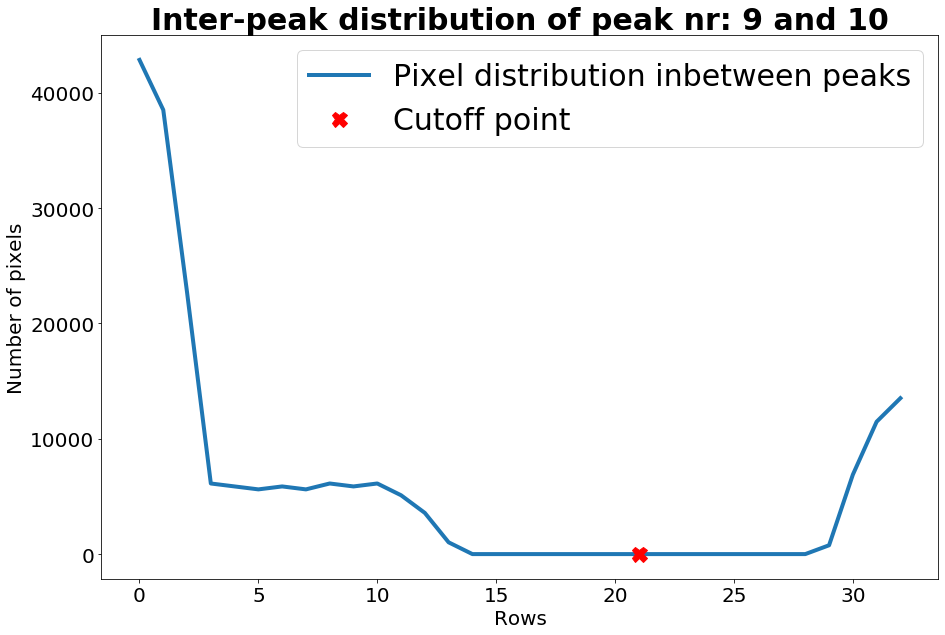} 
        \caption{An inter-peak interval where for every included row the corresponding amount of pixels is plotted. The median of all minima is marked (cutoff-point).}
        \label{fig:mesh05}
    \end{minipage}
\end{figure}

To do so, the image is mapped onto a one-dimensional vector of the length of all rows, each entry representing the number of pixels of that row (Figure \ref{fig:mesh04}). Performing this transformation yields a pixel distribution over the rows of the image. Peaks, or high amounts of pixel values, indicate the location of a signal whereas no pixels give insight that a gap had been found between signals. In order to find the perfect location for separation, each peak in the distribution has to be found with a restriction on a minimum distance to the next pixel peak. Without this constraint, too many pixel peaks are found. As our objective is to find the signals, adding the constraint results in more control.
\newline
Next, each interval between the found peaks is investigated. The minimal values in these inter-peak intervals serve as good candidates. The median-coordinate of all unearthed inter-peak minima is taken (Figure \ref{fig:mesh05})\footnote{The idea of this process was inspired by \cite[p.39f]{Fontanarava2019} and \cite{7868769} although a few adjustments have been made.}. 
\newline

The described process is iterated until every cutoff-coordinate is found. With the intention of further ensuring all signals are captured by this process, the interval from the last peak to the end of the picture is examined. If the encapsulated area has enough pixels inside (more than a certain threshold) another signal is found that would have been left out otherwise. \newline 
Type 1 ECGs show a different structure to the Type 2 and Type 3 ECGs. In the upper part of the images, one row consists of four recordings each from different channels. These have to be further separated to completely isolate every signal. The most satisfying results yields the cropping of each signal by finding the coordinates that are fully enclosing each signal. Luckily, for every Type 1 ECG image, these coordinates are the same such that a look-up table is established to provide a good cropping procedure. 	
Subsequently, the following step is to clean images of Type 2 and Type 3 from leftover noise, i.e., grid residuals. Again, the aforementioned finding-contour-with-threshold-procedure does justice to this task, assuring a signal that is as clean as possible at this stage. 	\newline 

Finally, each cropped signal is investigated by itself. A column-wise scan is executed where the extracted signal-coordinates are averaged over each column to obtain the inner signal line\footnote{As it is performed similarly by Ravichandran et al. \cite{Ravichandran2013}.}. However, these coordinates give no inside into the real time-voltage values. So, the coordinates have to be converted.

\subsection{Coordinate Conversion}
The goal of this step is to translate the pixel coordinate into time-voltage information. The previously removed background grid play an important part during this agenda. It gives inside into at what speed (25mm/s) and with how much voltage (10mV/s) the ECG was recorded. Using the information of how many pixels make up a big square of the background grid provides a good estimate for the pixel-coordinate to time-voltage-coordinate mapping. Of course, another distinction between the different types of ECG images has to be considered as the different types possess different kinds of grid resolutions.

\subsection{Upsampling}
To increase the number of sampling points and hence the amount of data obtained from the digitizing process, an upsampling of the extracted signal is conducted.\newline

A sample is nothing more than a number specifying the position of the signal\footnote{Analogous to sampling digital sound signals \cite{MDFTWEB07} \url{https://ccrma.stanford.edu/~jos/mdft/Introduction_Sampling.html}[date of last access: 19.12.2019].} and hence, the value the signal has at the time of measurement (sample-point). Upsampling has the aim of increasing the number of sampling-points and thus adding quantity to the data. This is achieved by a so-called \textit{zero-padding} of the signal in its frequency domain. 
Zero-padding can be understand as \say{extending a signal [..] with zeros} \cite{MDFTWEB07}\footnote{\url{https://ccrma.stanford.edu/~jos/mdft/Zero_Padding.html}[date of last access: 19.12.2019].}.\newline

First, the signal is transformed into the frequency domain as it lies naturally in the time domain. After all, it is a voltage signal recorded over time. The Discrete Fourier Transform (DFT) does justice to this operation since \say{the Fourier transform converts a signal that depends on time into
a representation that depends on frequency \cite[p.39]{Mueller2015}\footnote{The entire DFT Theorem and its derivation will be too much for the scope of this thesis. For the sake of completeness, it will be referred to \cite{Mueller2015} and \cite{MDFTWEB07}.}}.
\newline
Following the DFT, the signal is zero-padded. Zero-padding is defined by Smith \cite{MDFTWEB07}\footnote{\url{https://ccrma.stanford.edu/~jos/mdft/Zero_Padding.html}[date of last access: 19.12.2019].}:

\smallskip
\begin{figure}[H]
\begin{align}
ZEROPAD_{M,m}(x) \triangleq \begin{cases} x(m) & \text{$|m| < N/2$} \\ 0 & \text{otherwise} \end{cases}
\label{eq:03}
\end{align}
\end{figure}

where $N$ is the old length of the signal, $M$ is the new length of the signal and $m = 0, \pm 1,, \pm 2, \dots, \pm M_h$ with $M_h \triangleq (M-1)/2$ for $M$ odd, and $M/2-1$ for $M$ even. \newline

The size of the zero-padding is determined by a factor $L$ (which was chosen to be $L=8$) that also determines the upsampling factor in the time domain. Now, $N\left(L-1\right)$ zeros are inserted at exactly half of the frequency length in the frequency domain corresponding to the folding frequency (aka. half the sampling rate \cite{MDFTWEB07}\footnote{\url{https://ccrma.stanford.edu/~jos/mdft/footnode.html##foot20087}[date of last access: 06.01.2020] (listed as footnote 7.22).}).
\newline
As a final step, an inverse Fourier Transform \cite{MDFTWEB07}\footnote{\url{https://ccrma.stanford.edu/~jos/mdft/Inverse_DFT.html}[date of last access: 06.01.2020].} (IDFT) is used to map the sequence back to the time domain.\newline

The answer of how a zero-padding in the frequency domain is responsible for upsamling in the time domain is given by Smith \cite{MDFTWEB07}\footnote{\url{https://ccrma.stanford.edu/~jos/mdft/Periodic_Interpolation_Spectral_Zero.html}[date of last access: 19.12.2019].}, who states that \say{zero-padding in the \textit{frequency domain} corresponds to \textit{periodic interpolation in the time domain}} and further postulates the  following theorem:

\begin{figure}[H]
\begin{center}
\begin{align}
PERINTERP_L \triangleq IDFT(ZEROPAD_LN(X))
\label{eq:02}
\end{align}
$\forall x \in \mathbb{C} \text{ any any integer } L \geq 1$
\end{center}
\end{figure}

According to the theorem, a periodic interpolation is equivalent to a zero-padding of the signal in its frequency domain and subsequently transforming to the time domain. \newline

As the outcome of the upsampling procedure, a signal is yielded with an increased sampling rate and thus more data to investigate in subsequent steps. The reason for choosing this method for resampling becomes clear when the result is compared with other interpolation techniques. 

\begin{figure}[H]
\includegraphics[width=\linewidth]{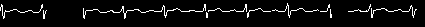}
\caption{Missing pixels}
\label{fig:missingpxl}
\end{figure}

\begin{figure}[H]
\includegraphics[width=0.6\paperwidth, height=0.25\paperheight]{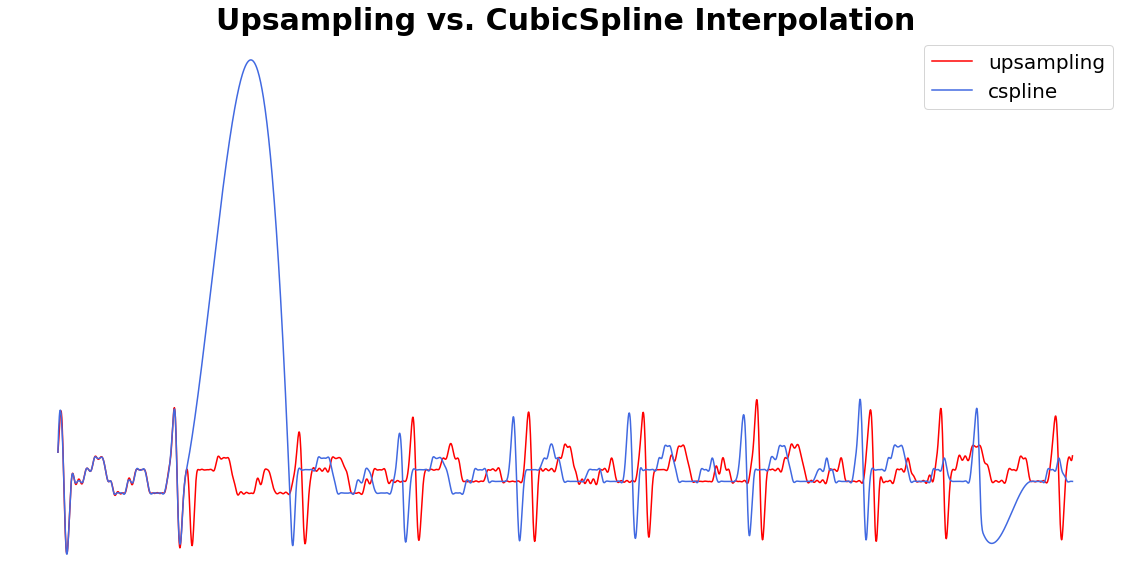}
\caption{Upsampling vs. Cubic Spline Interpolation}
\label{fig:upvsintpol}
\end{figure}

Figure \ref{fig:upvsintpol} shows an example of an interpolation process that led to an overshoot of the reconstructing sequence. A Cubic Spline interpolation is tested. Cubic spline interpolation is the smooth approximation of the underlying function by a piecewise polynomial\footnote{The entire theory of cubic spline interpolation is outside of the scope of this thesis. Yet, for the sake of completeness it is referred to \cite{Burger2015},\cite{Parker1983}.}.
\newline
The overshoot occurred from time to time, whenever there had been missing pixel values due to a too exhaustive cleansing procedure, see Figure \ref{fig:missingpxl}. Thus, it is preferred to take the noisy upsampled signal over the smoother interpolated signal\footnote{It is important to take in mind that the goal will be to feed the extracted data into an LSTM neural network which showed to be effective with noisy data, see \cite{Bizopoulos2019}.}. 

\subsection{Outcome}
The outcome of the upsampling procedure, and also previous processing steps, are several time-series of voltage data each representing one lead. One vector will be created for each channel and it will be subsequently stored in a data frame.

\begin{figure}[H]
    \centering
    \includegraphics[page=2, scale=0.45]{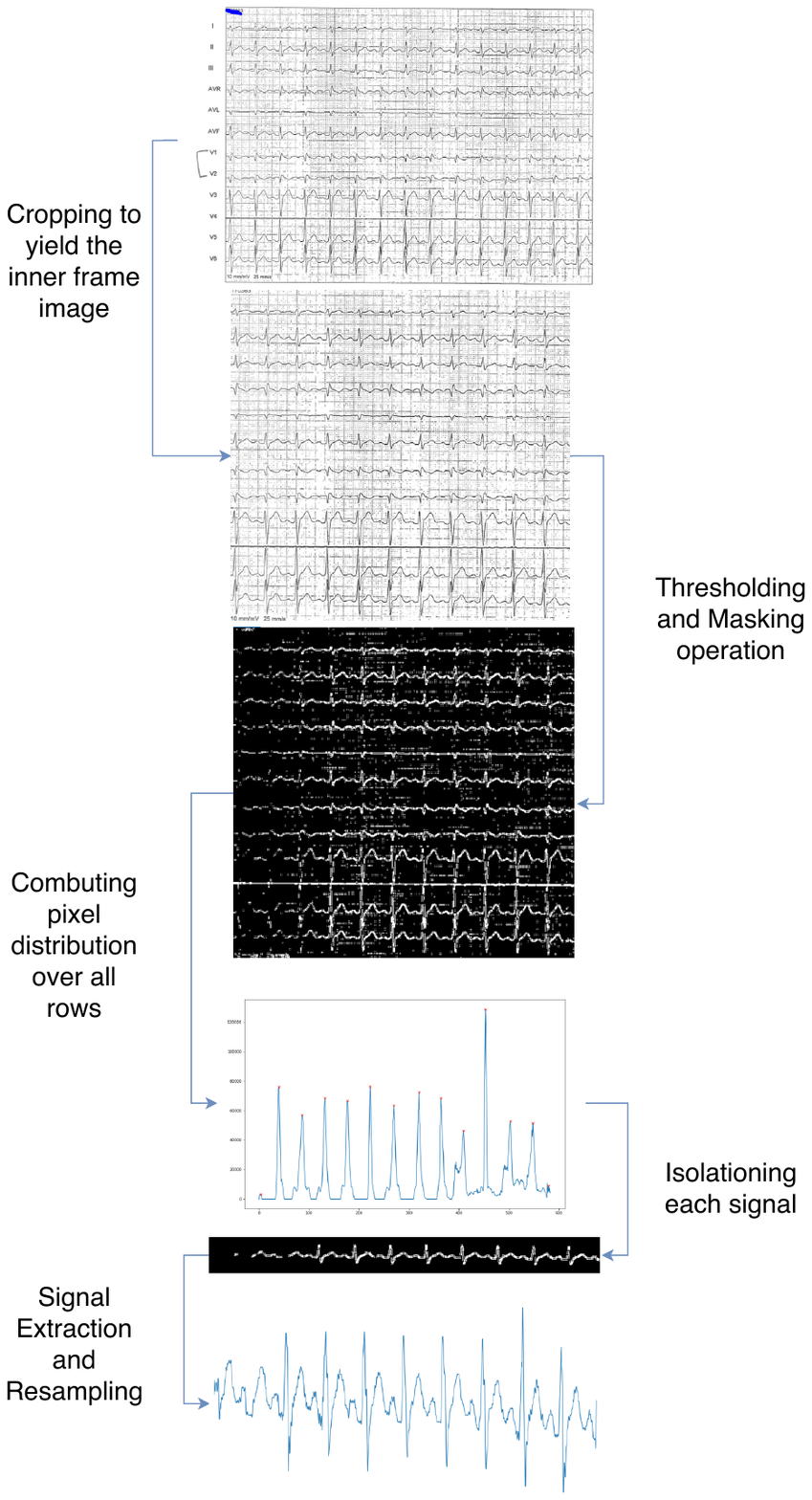}
    \caption{Type 2 ECG digitization process.}
    \centering
\end{figure}

\section{Results}
The aforementioned pipeline takes in scans of ECGs and digitizes them in a manner that each lead will be reconstructed via an approximation of the signal. Hereby, the integrity of each lead is preserved in most cases. However, the isolation of a signal is sometimes faulty due to strongly overlapping signals, that could not be fully separated, characters that were not fully removable since they are overlapping with the signal or, as in the case of Type 2 images, lost signal parts. 

\begin{figure}[H]
    \begin{subfigure}{\linewidth}
    \includegraphics[width=\linewidth]{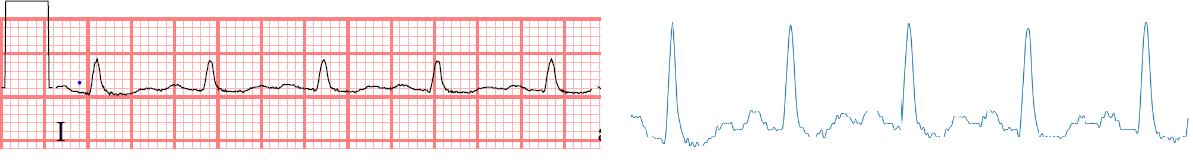}
    \caption{}
    \end{subfigure}\par\medskip
    \begin{subfigure}{\linewidth}
    \includegraphics[width=\linewidth]{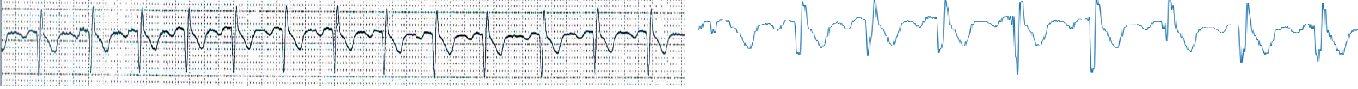}
    \caption{}
    \end{subfigure}\par\medskip
    \begin{subfigure}{\linewidth}
    \includegraphics[width=\linewidth]{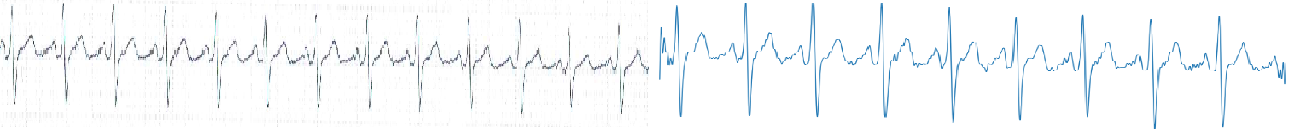}
    \caption{}
    \end{subfigure}
    \caption{The successfully reconstructed signal of (a)  lead 1 of a Type 1 ECG, (b) lead aVR of a Type 2 ECG and (c) lead 1 of a Type 3 ECG. Note: the line of (a) appears to be fading. This is due to the fact that the two images have a different resolution and (a) is slightly distorted.}
    \label{fig:mesh2}
    \centering
\end{figure}

Figure \ref{fig:mesh2} depicts an example of a successful recovery of a recorded ECG lead for each different type. On the left side, the original scanned ECG signals depicted and on the right side, the recovered signals are presented. All three image types are correctly captured by the pipeline as their features are preserved and the waveform is identical. 

\begin{figure}[H]
    \begin{subfigure}{\linewidth}
    \includegraphics[width=\linewidth]{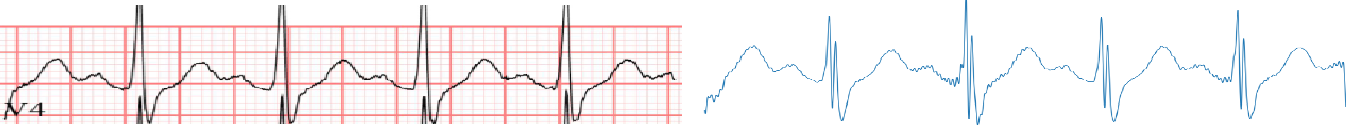}
    \caption{}
    \end{subfigure}\par\medskip
    \begin{subfigure}{\linewidth}
    \includegraphics[width=0.95\linewidth]{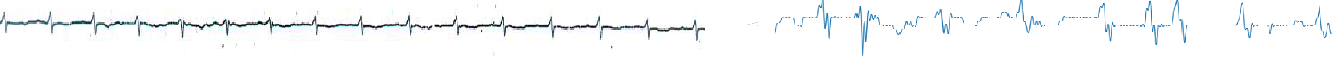}
    \caption{}
    \end{subfigure}\par\medskip
    \begin{subfigure}{\linewidth}
    \includegraphics[width=\linewidth]{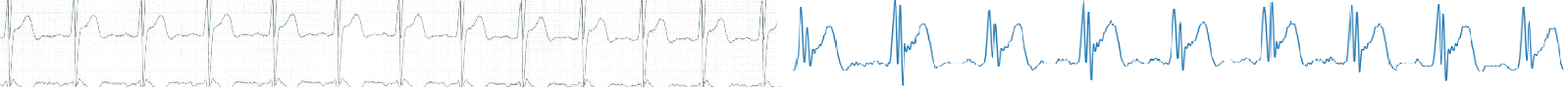}
    \caption{}
    \end{subfigure}
    \caption{Retrieved signals that contain faults from (a) a Type 1 V4 lead, (b) a Type 2 AVL lead and (c) a Type 3 V4 lead.}
    \label{fig:mesh3}
\centering
\end{figure}

Figure \ref{fig:mesh3} illustrates the example of a faulty recovery procedure. In the case of (a) and (c), the peaks of the signal are deformed by another signals' peaks that could not be separated from the crop. Hence, the peak has a sudden drop where the two signals are overlapping in the column. Plus, images of Type 1, sometimes were additionally affected by leftover characters that could not be fully removed in the processing steps and thus also affecting the accuracy of the recovered signal (see (a)). \newline
As for Type 2 images, due to the lack of quality, it was often not possible for the pipeline to distinguish background and foreground such that parts of the signal were erased or falsified by residuals that were too strongly intertwined with the signal to have them removed in one of the filtering procedures. Hence, the restoration does not capture the original signal. 

\section{Discussion}
The proposed pipeline provides a framework for establishing a fully automated digitization process. A distinction between different types of ECG, which the pipeline had to face during the agenda, resulting in a workflow that united several steps but with diverging elements depending on what type is currently processed.
\newline
The quality of the recovered signals is repeatedly captured in a remarkable quality when the signals are isolated and not too strongly enveloped by the background grid. Here, the quality does not differ from the results presented in the literature. Likewise, the quality of the result drops as soon as the signals are strongly overlapping with each other and a clear separation, by taking only the height-coordinates into account, is not possible anymore, or if the background grid strongly interferes with the signal. The success depends further on the quality of the scanner used and on the tidiness of the action itself. \newline
Of course, the presented pipeline appears to be tailored specifically for the three distinct types of ECGs faced. An introduction of a new form of paper ECG would require more inspection and an additional parameter search. However, this procedure is one of the first to investigate distinct types (Fontanarava \cite{Fontanarava2019} uses two distinct types) and attempting to unify them within the same procedure. \newline
Future work lies in the task of implementing an improved signal separation. One way could be to use CNNs as presented by Fontanavara \cite{Fontanarava2019} that captures the main components of the signal such that it can be further extracted. For this thesis, it was decided to persist with a human-engineered pipeline.
Additionally, an improvement could be made by consulting already digitized ECG signals and their scanned paper versions in order to compare the results of the aforementioned pipeline to create an optimization process by minimizing a metric between the original signal and its reconstruction. \newline

Overall, the quality of the recovered data is sufficient to be used for the second part of this thesis, in which we will use the data to construct a classifier that is able to distinguish BrS positive and BrS negative patients based on there ECG.

\section{Conclusion}
In this part of treatise, a method was proposed to digitize and recreate ECG signals previously only accessible in paper form. This method encompasses three different types of ECGs such that all signals can be retrieved from each type. Processing steps differed depending on the type and consisted sometimes of an extra step and sometimes only of a difference in the parameter settings. The results of the process have been highlighted, strengths and weaknesses were disclosed.  Furthermore, an outlook in future work has been introduced which could lead to improved workflow. \newline
The aim of digitization has been reached and the extracted data promotes a basis for further investigation. 

\chapter{Part II: Learning to Diagnose}
\setcounter{page}{19}
\section{Introduction}
In the second part of the thesis, we propose the classifier that is built based on the data gathered during Part I and an additional database.
\newline

We start by  stating the problem and the objective of this part. Thereafter, we acquaint ourselves with various applications of machine learning in cardiology. Then, we describe the fundamentals that underlie the LSTM network. In the subsequent chapter, the data is presented together with necessary preprocessing steps. Following the data chapter, we propose the model together with all the important settings that have to be made. Afterward, we investigate the performance of the model in an evaluation procedure and discuss the results.

\subsection{Problem Statement}
During a lifetime, the heart is neither supposed to rest nor stop, otherwise the effects might be fatal. However, different kinds of impairments and diseases can impose serious problems on the living organism. Cardiovascular diseases (CVDs) are the number one cause of death in a global spectrum. In 2016, the World Health Organization (WHO) reported that  17.9 million people died from CVDs which resulted in 31.9\% of all global deaths, infarcts and strokes were representing the main causes of death\footnote{\url{https://www.who.int/en/news-room/fact-sheets/detail/cardiovascular-diseases-(cvds)}[date of last access: 21.01.2020].}.
\newline

Sudden cardiac arrests (SCAs) are a grave case of CVDs that lead to a swift loss of the cardiac functions and subsequently pose an imminent threat to human life \cite{Albert2018}. If the outcome is fatal, then the event is referred to as sudden cardiac death (SCD) \cite{Albert2018}. 
\newline
The Brugada Syndrome (BrS) is a special type of SCA that often results in an SCD. BrS is diagnosed by analyzing the ECG since it is characterized by abnormalities in the signal \cite{BRUGADA2006S115}. Ever since its first description in 1992 by Pedro and Josep Brugada \cite{Brugada1992}, the syndrome is intensely studied. BrS is associated with a high rate of sudden cardiac death (4\% of the total count) in patients with structurally normal hearts (up to 20\%) and appears to result mostly in an occasion of sudden death for men around the age of 41 years \cite{Charles2005}.

\newpage
\begin{figure}[H]
    \begin{subfigure}{\linewidth}
    \includegraphics[scale=1.5]{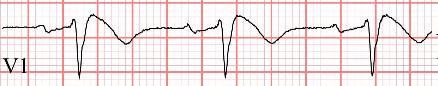}
    \centering
    \caption{}
    \end{subfigure}\par\medskip
    \begin{subfigure}{\linewidth}
    \includegraphics[scale=0.8]{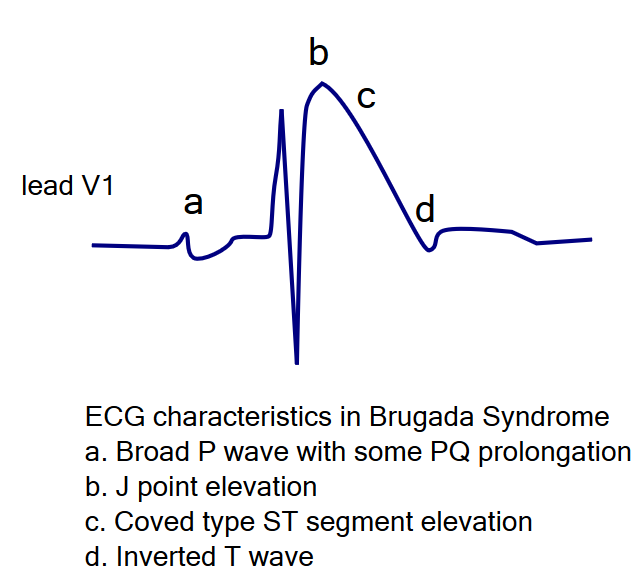}
    \centering
    \caption{}
    \end{subfigure}
    \caption{The typical BrS-pattern in a V1 lead. (a) taken from the own data base and (b) taken from \url{https://en.ecgpedia.org/images/f/f1/Brugada_ecg_characteristics.svg} [date of last access 06.02.2020].}
    \label{fig:ecgBrS}
    \centering
\end{figure}

The only diagnostic case is characterized by an accentuation of the J wave found in the right precordial leads (V1, V2), which results in an ST-segment elevation that is often followed by a negative T-wave \cite{Antzelevitch2012},\cite{Brugada1992} (see Figure \ref{fig:ecgBrS}). This behavior can be either spontaneous or drug-induced \cite{Wilde2002}. There exist also other criteria that hint towards BrS (see \cite{Wilde2002} or \cite{Waks2017}). However, these criteria \say{are based on currently available data and that it is a work-in-progress that is awaiting confirmatory [...] clinical data \cite{Wilde2002}.}
\newline

In order to make precise decisions about the diagnostic procedure in BrS, new criteria have to be found to distinguish the cases. The call for a classifier - able to distinguish between BrS positive and negative subjects - cannot be ignored. Thus, the challenge will be faced and its course is presented to the reader throughout this second part of the thesis. 

\subsection{Objective}
During this part, the construction process of a binary classifier that can distinguish and correctly classify BrS positive from BrS negative patients solely based on their presented ECG data is displayed. Moreover, the classifier is built on the extracted time-series data from Part I. It is hoped that this classifier will be able to give an additional opinion when it comes to the diagnosis of the rare heart disease. \newline
Everything described here is implemented using Python \cite{Python3} 3.7.3 and the libraries used are Pandas \cite{Pandas}, Numpy \cite{Numpy}, Scikit-Learn \cite{scikit-learn}, PyTorch \cite{PYTORCH}, Matplotlib \cite{matplotlib} and Seaborn \cite{Seaborn}.

\section{Related Work}
Every day, a vast amount of data is generated in the medical field. It stems from vital signs of patients, genetics or diseases and it is collected and processed to evaluate and improve methods or for the creation of new methods. 
\newline 
Machine Learning provides the right tools to analyze the tremendous amount of information. The generation of statistical models, that can classify substructures within the data or make predictions about future states, and the subsequent evaluation of their performances leads to more insight into the field. Sophisticated methods thus have the rightful potential to become an accomplice to the human nurses and doctors, improving the quality of medical aid.
\newline 
Since, for this thesis, we target a specific heart disease, it is important to inspect the role of machine learning in cardiology focusing on methods that rely on ECG data. In the following, an overview of this field is presented and then a special kind of method, the LSTM, is introduced to lay a foundation of understanding the subsequent chapters.

\subsection{Machine Learning in Cardiology}
The medical field provides the perfect playground for testing various machine learning techniques as there is a constant generation of data. However, capturing the data does not always occur cleanly but rather it can be faulty due to measuring errors. Therefore, working with medical data provides a good opportunity to test techniques on real-life practical data and not only on artificially constructed data sets. ECG signals constitute an example of the data generated in the medical field and more precisely in cardiology. The periodic signal serves as an indicator of the vital status of the human heart and its features give account over different conditions. 
\newline
Deep Learning techniques are well-fitting candidates when it comes to the analysis and feature detection within ECG signals \cite{Bizopoulos2019}. Especially, the Recurrent Neural Network (RNN) is a suited device to find structure in temporal data \cite{Bizopoulos2019}, \cite[p.273]{Aggarwal2018} and provided good results in arrhythmia detection \cite{Lipton2015}, \cite{OH2018278} (the presented cases are based on the LSTM, an improved successor of the vanilla RNN model).  
\newline 
The LSTM showed prosper results in related fields that use time-series data \cite{Liu2014},\cite{Siami-Namini2018}. Recent propositions to apply the LSTM networks onto medical data resulted in promising outcomes \cite{Lipton2015},\cite{Han-Gyu2017}, \cite{OH2018278}. Lipton et al. \cite{Lipton2015} were able to construct a multi-label LSTM classifier that is fed with raw medical data and not only predicts missing data but also gives a diagnosis via the classification label in the end. While Oh et al. \cite{OH2018278} used a model that combined a CNN with an LSTM to diagnose arrhythmias in ECGs. 
\newline Before we come to the presentation of the data and subsequently to the LSTM-architecture that is used to classify ECG signals in the case of the BrS, we first have to take a look at the fundamentals to fully understand the given structure. 

\subsection{LSTM Fundamentals}
By using an Artificial Neural Network (ANN), the goal is always to construct a function approximator that, depending on the problem, can receive data as input and return, i.e., an assigned label or a replicated sequence.
\newline
The network consists of an input layer with one unit per input-feature, several optional hidden layers with arbitrary size and an output layer, where the amount of units depends on the specific problem\footnote{To illustrate, one output unit is sufficient for a binary classification problem in which the positive classification is labeled as $1$ and the negative one as $0$. The output of the ANN is then usually a probability between $0-1$.}. Each layer is connected to its previous with weights (the weight vector is denoted as $\vec{\theta}$). The entire output of the previous layer arrives at a unit of the current layer as a weighted sum, which in turn is used to feed an activation function $\phi$, that is often nonlinear such that the output is again only a single scalar \cite[p.9]{Johnson2019}. An arbitrary layer, $h_j$, can be seen as a function of the layer that proceeded it $h_{j-1}$ \cite[p.193f]{Goodfellow-et-al-2016}. Therefore, our function approximator $\hat{f}$ can be seen as a chain of functions \cite[p.164]{Goodfellow-et-al-2016}. In the end, a mapping is learned with parameters $\theta$ \say{that result in the best function approximation \cite[p.164]{Goodfellow-et-al-2016}}.
\newline
In some cases, the stream can have also backward directions and thus loops. Those networks are then called recurrent neural networks \cite{nielsenneural}.
\smallskip

\begin{equation}
    \hat{y} = \hat{f}(\vec{x},\vec{\theta})
\end{equation}
\smallskip

During the learning phase, the output, $\hat{y}$, of the function approximator is compared to a target value\footnote{Given a supervised learning task.}, $y$, with whom the loss (or cost) function is calculated. To become a good function approximator, the network has to minimize the loss function (the calculated difference between $y$ and $\hat{y}$). It does so, by updating its weights $\theta$, that have been used to compute the final output, using gradient descent optimization, which means that the network is making small updates every iteration in direction of the steepest descent of the loss function \cite{BackProp95},\cite[p.80ff]{Goodfellow-et-al-2016}. Throughout this procedure, the error is backpropagated, meaning that the error of the output is used to calculate inner-network-errors to affect all weights of the network\footnote{The general backpropagation algorithm is described in \cite{BackProp95}, \cite{Goodfellow-et-al-2016}.}.
\newline

The LSTM\footnote{The architecture presented here is the one presented by Gers et al. \cite{Gers} with the additional Forget Gate.}, developed by Hochreiter et al. \cite{LSTM97}, is a more sophisticated neural network that can pass information over time, due to its recurrent nature, and take sequences as an input. Its most remarkable feature is a memory cell that \say{can maintain its state over time \cite[p.1]{Greff15}}, which is passed from time step to time step.
\newline

A forward pass of the LSTM is described as follows \cite{Gers},\cite{Olah15},\cite{LSTMWIKI}:
\newline 
Input $\Vec{x}_t$, that is fed into the network at time step $t$, will be first passed through several gates that \say{regulate the information flow into and out of the cell \cite[p.1]{Greff15}}.
The first one is the forget gate, $f_t$, that decides which information will be kept inside the memory block $C_{t-1}$ and which information will be discarded.
\newline
Next, the relevant data is determined that will be used to update the memory cell. This process is split into two parts. During the first part, it is resolved which values of $C_{t-1}$ (the cell state at time step $t-1$) will be updated bypassing the input through an input gate, $i_t$. Then, new candidate values are computed by guiding the input through a candidate gate, $\Tilde{C}_t$.
\newline
Subsequently, the new cell state, $C_t,$ is determined by taking into account which values are to be forgotten determined by $f_t$ and which candidate values $\Tilde{C}_t$ will replace which old values determined by $i_t$.
\newline
Lastly, the output, $h_t$, is determined by passing the cell state through a $tanh$-function and taking into account the relevant output elements that were determined by an output gate, $o_t$.

\begin{figure}[H]
\begin{align}
& f_t = \sigma(\Vec{W}_f \ \Vec{x}_t + \Vec{U}_f \ h_{t-1} + \vec{b}_f)\\
& i_t = \sigma(\Vec{W}_i \ \Vec{x}_t + \Vec{U}_i \ h_{t-1} + \vec{b}_i)\\
& o_t = \sigma(\Vec{W}_o \ \Vec{x}_t + \Vec{U}_o \ h_{t-1} + \vec{b}_o) \\
& \Tilde{C}_t =  \mathrm{g}(\Vec{W}_{\Tilde{C}} \ \Vec{x}_t + \Vec{U}_{\Tilde{C}} \ h_{t-1} + \vec{b}_{\Tilde{C}})\\
& C_t = f_t \odot C_{t-1} + i_t \odot \Tilde{C}_t \\
& h_t = o_t \odot \mathrm{h}(C_t)
\end{align}
\caption{Forward pass of an LSTM network for time step $t$. \newline 
Let $x_t \in (x_1, \dots, x_T)$ denote the input at time step $t$, $\Vec{W}$ the input weights, $\Vec{U}$ the weights of the hidden unit and $\vec{b}$ the respective biases. Further, $\odot$ denotes the element-wise product, $\sigma(x) = \frac{1}{1-e^{-x}}$ the logistic sigmoid and $\mathrm{h}(x)=\mathrm{g}(x)=tanh(x)=\frac{sinh(x)}{cosh(x)}=\frac{e^x -e^{-x}}{e^{x} + e^{-x}}$ the hyperbolic tangent.}
 
\label{fig:eq04}
\end{figure}

Once the input is fed through the network at the respective time step $t$, it is time to update the network's parameters\newline
$\theta = (\Vec{W}_f, \Vec{W}_i, \Vec{W}_h, \Vec{W}_{\Tilde{C}}, \Vec{U}_f, \Vec{U}_i, \Vec{U}_h, \Vec{U}_{\Tilde{C}}, \Vec{b}_f, \Vec{b}_i, \Vec{b}_h, \Vec{b}_{\Tilde{C}})$. \newline
To do so, the error between the target output at time $t$, $y_t$, and the network output, $h_t$, is calculated with the aim to minimize it during the learning procedure by using a loss function, $l$. 
\begin{equation}
    Error_t = l(y_t-h(x_t))
\end{equation}

After defining the error term, we can compute the error of an arbitrary gating unit $u_t \in (f_t, i_t, o_t)$ at time step $t$ by taking the partial derivative of the error w.r.t. the specific gate and considering the chain rule\footnote{In Leibniz notation: $\frac{\partial z}{\partial x}=\frac{\partial z}{\partial y}\frac{\partial y}{\partial x}$, whenever a variable $z$ is dependent on a variable $y$, which itself is dependent on a variable $x$ \cite{ChainRuleWiki} and we are eager to calculate the derivative of variable $z$ w.r.t. $x$.}.
\vspace{0.3cm}
\begin{equation}
    \frac{\partial Error_t}{\partial u_t} = \frac{\partial Error_t}{\partial h_t}\frac{\partial h_t}{\partial u_t}
    \vspace{0.3cm}
\end{equation}

Now, our goal will be to update the weights based on their impact on the calculated error. Therefore, we will search for weights that minimize the error, and thus minimizing the loss function, by calculating $\frac{\partial Error_t}{\partial \theta_t}$.

\begin{figure}[H]
\begin{align}
& \delta h_t = \Delta_t + \Delta h_t \\
& \delta C_t = \delta h_t \odot o_t \odot \mathrm{h}'(C_t) + \delta C_{t+1} \odot f_{t+1} \\
& \delta\Tilde{C} = \delta C_t \odot i_t \odot \mathrm{g}'(\Tilde{C}_t) \\
& \delta o_t = \delta h_t \odot \mathrm{h}(C_t) \odot \sigma '(o_t) \\
& \delta i_t = \delta C_t \odot \Tilde{C}_t \odot \sigma '(i_t) \\
& \delta f_t = \delta C_t \odot C_{t-1} \odot \sigma '(f_t) \\
& \delta x_t = \sum_{u \in (f_t, i_t, o_t)}\vec{W}_u^T \delta u_t \\
& \Delta h_{t-1} = \sum_{u \in (f_t, i_t, o_t)}\vec{U}_u^T \delta u_t
\end{align}
\caption{Backwards error flow}
\label{fig:eq05}
\end{figure}

Figure \ref{fig:eq05} describes such a backward pass\footnote{The notation is a mix out of the named resources and own notation to stay consistent with the previously used notation.} \cite{LSTM97},\cite{Greff15},\cite{chen2016gentle}.
\newline
First, the error $\delta h_t$ of the output is calculated by using $\Delta_t = \frac{\partial E(\vec{x}_t)}{\partial h}$, the error of time step t, and $\Delta h_t = l(y_{t+1}, h_{t+1})$ (recursively defined in Equation \ref{fig:eq05}) the output difference \cite{LSTMBackpropMedium} computed at the subsequent time step $t+1$.
Next, the error $\delta C_t$ is computed as it has been shown by Hochreiter et al. \cite[p.27]{LSTM97} and Chen \cite[p.8]{chen2016gentle}.
\vspace{0.3cm}
\begin{equation}
    \delta C_t := \frac{\partial E_t}{\partial C_t} + \frac{\partial E_{t+1}}{\partial C_t} = \frac{\partial E_t}{\partial h}\frac{\partial h}{\partial C_t} + \frac{\partial E_{t+1}}{\partial h_{t+1}}\frac{\partial h_{t+1}}{\partial C_{t+1}}\frac{\partial C_{t+1}}{\partial C_t}
    \vspace{0.3cm}
\end{equation} 

This behavior is because the error is not only backpropagated by the error $E_t$ but also via the memory cell $C_t$. The equation given above can be now rewritten as 
\vspace{0.3cm}
\begin{equation}
    \delta C_t = \delta h \odot o_t \odot \mathrm{h}'(C_t) + \delta C_{t+1} \odot f_{t+1}
    \vspace{0.3cm}
\end{equation} 

considering the equations given in the forward pass (Figure \ref{fig:eq04}).
\newline 

The gate updates can be computed in a similar fashion\footnote{A full derivation of the equations is given in \cite{LSTM97} and \cite{chen2016gentle}.}. Note that $\delta x_t$, the error for the inputs, is only required when there exists an additional layer below the LSTM (i.e., a second LSTM layer) that requires training \cite{Greff15}.
\newline 

Finally, the parameters are updated 
\begin{align}
    &\delta \vec{W}_u = \sum_{t=0}^T \delta u_t \otimes \vec{x}_t \\
    & \delta \vec{U}_u = \sum_{t=0}^{T-1} \delta u_{t+1} \otimes \vec{x}_t \\
    & \delta \vec{b}_u = \sum_{t=0}^T \delta u_{t} 
\end{align}

where $\otimes$ denotes the outer product of two vectors and considering a stochastic gradient descent algorithm, with a learning rate of $\eta$, the final parameter update becomes
\medskip 

\begin{equation}
    \theta^{new} = \theta^{old} + \eta\ast\delta\theta^{old}.
\end{equation}
\medskip

The entire process is repeated for every time step until the sequence was fed in completely. 

\section{Data}
In the following, the positive and negative data is displayed as well as the necessary preparation steps. However, the data is not searched for features or normalized. The preparation of the raw data only covers for necessary steps to run the algorithm. We aim for a proof of concept rather than a state-of-the-art solution, since many factors, e.g., the available data, were not as progressed by the time of writing this thesis. 

\subsection{Positive Training Data}
We label positive examples as positive because they are composed of ECG signals that were taken from patients with a diagnosed BrS. Each example consists of a recorded signal derived from its respective lead.
Since digital signals were not available in the case of BrS positive patients, the learner is fully dependent on the ECG signals that were recovered during Part I of this thesis, namely the digitization process. Hereby, the layout of the three different types of images led to signals with varying length depending on whether they stem from Type 1, Type 2 or Type 3 ECGs. Furthermore, the quality also varies since some of the ECGs were not entirely recoverable as explained during Part I.
\newline
By the time of conducting the experiment, 30 positive training instances each consisting of 3000 data points are available.

\subsection{Negative Training Data}
Negative training data is negative in the sense that the recorded signals stem from healthy control patients. They were gathered from the PTB Database of Physionet \cite{bousseljot_kreiseler_schnabel_2009}, \cite{Physionet}. The database contains 549 records from 290 subjects and a mean age of 57,2. A detailed clinical summary of the patients is attached, however, this information is only used to locate the healthy control patients and the rest of the information is discarded. ECGs from this database have been recorded in 16 (I, II, III, aVR, aVL, aVF, V1, V2, V3, V4, V5, V6, VX, VY, VZ) different leads and digitized at 1000 samples per second. In total, the digital ECGs of 52 healthy control patients can be extracted and their data is used to build the model.
\newline
In total, 80 negative training examples are used with a respective length of 3000 per instance.

\subsection{Data Preparation}
For the experiment, we select to discard signals from leads other than V1 or V2, as they are the primary indicators for the BrS (see 3.1.1). Furthermore, three separate models, one for V1, one for V2 and one for V1 \& V2 (also referred to as "both" in the figures), are created to test them against each other. The data is randomly shuffled and then split into training ($70\%$), validation ($15\%$) and test data ($15\%$) responsible for training the model, testing hyperparameters and evaluating the final performance respectively. In the same moment, labels are assigned to each instance where \textit{1} indicates BrS positive and \textit{0} denotes BrS negative. \newline
Before entering the LSTM-system, the data sequence is further split into non-overlapping windows that consist of 500 data points each. This is done to further increase the amount of data available.
A final consideration has to be made as the two data sets (positive and negative instances) are unbalanced in their quantity. Thus, we decide to take action against this, as it would lead to difficulties when evaluating the performance of the classifier (this is also thematized in 3.5.1), by assigning weights during training to the training instances. Positive instances are given a weight of $3.67$, according to the ratio $\frac{\text{neg. \ instances}}{\text{pos. \ instances}}$ in the training data set, as it is suggested in the Pytorch documentation\footnote{\url{https://pytorch.org/docs/stable/nn.html##bcewithlogitsloss} [date of last access: 05.02.2020].}, and thus the loss computation is adjusted as if the dataset is equally balanced. 

\section{Method}
\begin{figure}[H]
    \begin{center}
    \includegraphics[scale=0.75]{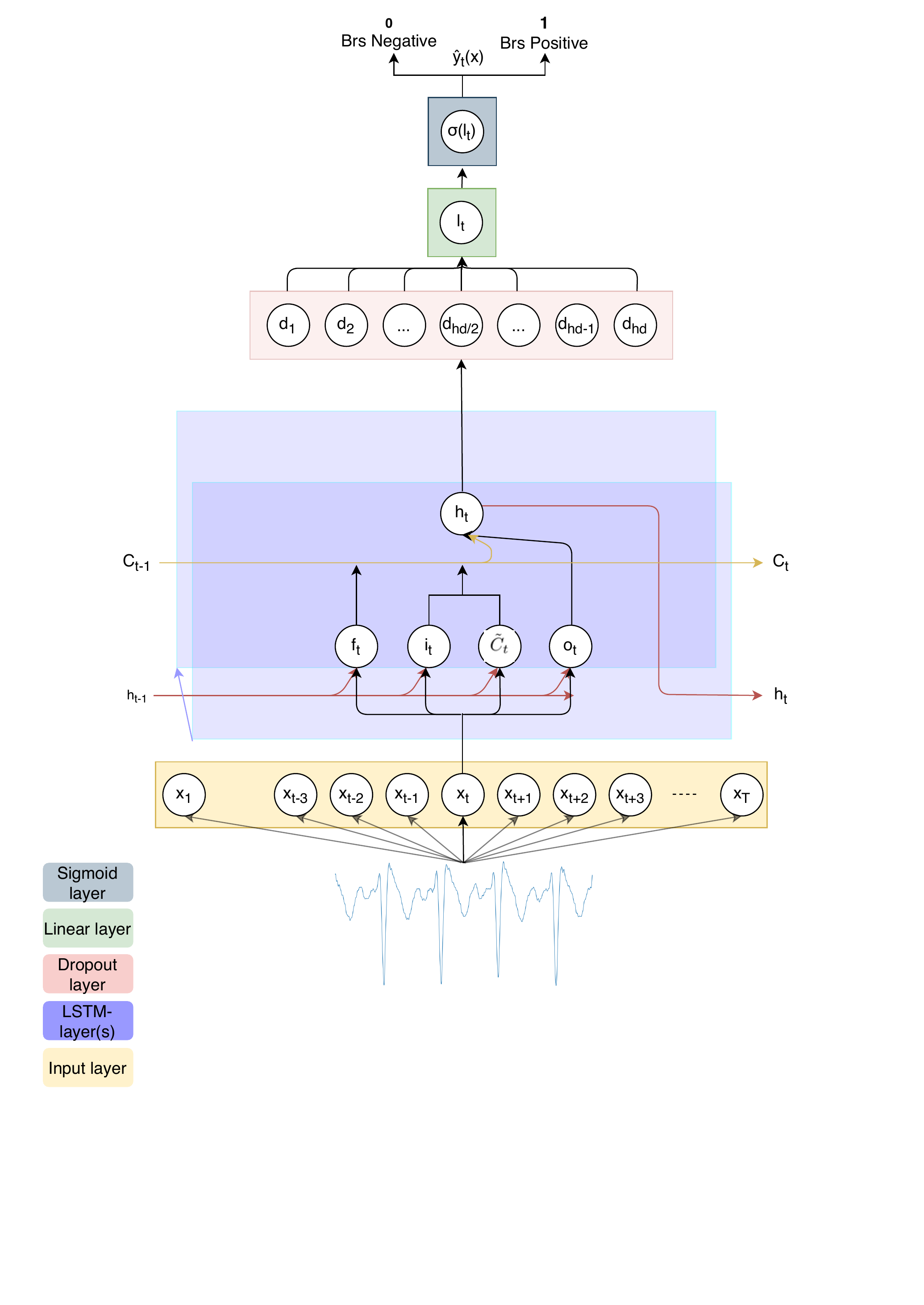}
    \caption{Architecture of the Neural Network used during the experiments}
    \centering
    \label{fig:LSTM}
    \end{center}
\end{figure}

The ANN that is used throughout the experiments is depicted in Figure \ref{fig:LSTM}. Input to this network is the raw ECG signal, either extracted via the pipeline described in Part I or taken from the database. The ECG signals constitute time-series data, where each data point represents the input of $x_t$ at time $t$. Inputs at time $t$ are forwarded into the LSTM-layer where they are subject to the forward pass computations that were described in 3.2.2.
\newline
We introduce a dropout layer (see 3.4.1), that randomly (chance of 25\%) omits neurons and is placed after the LSTM-layer to avoid overfitting to the small data set. Additionally, a linear layer that constitutes as a bottleneck and squishes the dimensionality down to one, the output dimension, is added after the dropout layer.
\newline
In the final step, the output of the network is passed through a sigmoid function $\sigma(x)=\frac{1}{1-e^{-x}}$ such that the final output $\hat{y} \in [0,1]$ constitutes for the probability of the ECG belonging to class $0$ (BrS negative) or class $1$ (BrS positive). Subsequently, the loss is calculated by using a loss function which is specified in its subsection. Then, the network's parameters are updated in accordance with the loss function operating the backpropagation described in the second half of 3.2.2. 
\newline
In this chapter, the model settings such as the chosen loss function, optimizer and dropout are highlighted. However, parameter settings, such as the number of epochs or the chosen learning rate, are part of the results chapter since they are found after a parameter search according to the criteria of measurement, which are to be presented in the following chapter.

\subsection{Dropout}
Neural networks tend to have many different parameters that allow for a variety of different parameter settings. A lot of which can lead to a good result in the training procedure. However, most of these different configurations will be worse performance-wise when being confronted with the test data as compared to being confronted with the training data \cite{hinton2012improving}\footnote{A phenomenon called overfitting \cite[p.364]{hastie_09_elements-of.statistical-learning}.}. This is because some neurons within the network depend too much on other neurons of the network \cite{hinton2012improving}. Dropout is used to regularize a neural network during the learning procedure and thus reduce its chance of overfitting to the data \cite[p.189f]{Aggarwal2018}. Dropout works by randomly omitting a subsample of all of the network's neurons by a predefined chance during the presentation of each training example \cite{hinton2012improving}. The idea is that neurons will stop being too reliable on other neurons and thus, stop learning co-dependencies \cite{hinton2012improving}.
\newline

The here presented network possesses a dropout-layer that follows right after the output of the LSTM-layer. Dropout occurs with a chance of $25\%=0.25$. Additionally, a dropout within the LSTM-layer is implemented  with a chance of $0.25$, if the number of stacked LSTM-layers exceeds $1$.

\subsection{Cost Function}
In a supervised learning task, the learner (in our case the LSTM model), will generate outputs based on the learned function $\hat{f}(X)$. Given the inputs $X = x_1, x_2, \dots, x_N$, outputs are produced in the form $\hat{f}(x_i)$. During the training process, the learner is able to \say{modify its input/output relationship $\hat{f}$ in response to differences, $y_i-\hat{f}(x_i)$, between the original and generated outputs \cite[p.48]{hastie_09_elements-of.statistical-learning}}. By an iterative adjusting procedure, it is hoped that $\hat{f}$ becomes close to $f$, the real underlying function. 
\newline
Imagine that the target outputs, $y_i$, stem from a probability distribution, $Y$, such that $Y = f(X) + \epsilon$, where $X$ describes the probability distribution of the training instances and $\epsilon$ denotes a small noise term\footnote{The usage of $\epsilon$ is justified as \say{for most systems the input-output pairs $(X,Y)$ will not have a deterministic relationship $Y = f(X)$ \cite[p.47]{hastie_09_elements-of.statistical-learning}}.}. Now, together with our own approximator, $\hat{f}$, and the training instances we can construct the probability distribution $\hat{Y} = \hat{f}(X)$. 
\newline
The only thing that is left for us is to find a way of how we can compare the two distributions such that they will be as close as possible by the end of the training.
\smallskip

\begin{align}
    & H(p,q) = H(p) + D(p||q) \\
    & = -\sum_{x \in X}p(x) \ log \ q(x) 
\end{align}
\smallskip

The equation given above is the cross-entropy. It is composed of the entropy of the variable $X$ together with the relative entropy (or Kullback Leibler distance). The Entropy of a variable gives account over the uncertainty of its outcome, whereas the relative entropy is a \say{measure of the distance between two distributions \cite[p.18]{InformationTheory}}. The latter is used when two probability distributions, $P(X)$ and $Q(X)$, exist over the same variable $X$ \cite[p.72]{Goodfellow-et-al-2016} and their similarity are measured. \newline
Since the goal for our generated distribution, $\hat{Y}$, is to become as close as possible to the real target distribution $Y$, we strive to minimize the Kullback Leibler distance and thus equivalently minimize the cross-entropy \cite[p.130]{Goodfellow-et-al-2016} (as both terms, entropy and relative entropy, are defined positively\footnote{$H(X) = -\sum_{x \in X} p(x) \ log \ p(x)$  and $D(p||q) = \sum_{x\in X} p(x) log\frac{p(x)}{q(x)}$ defined in \cite{InformationTheory}.}). Therefore, the information-theoretical definition gives a nice intuition of what is happening when the cross-entropy is minimized, namely that the dissimilarity between the two distributions $\hat{Y}$ and $Y$ is minimized. 
\newline

Remarkably, while minimizing the cross-entropy and thus finding the optimal $\theta$, we also maximize the maximum likelihood for our parameters\footnote{\say{The principle of maximum likelihood assumes that the most reasonable values for $\theta$ are those for which the probability of the observed sample is largest \cite[p.50]{hastie_09_elements-of.statistical-learning}.}} \cite[p.31f]{hastie_09_elements-of.statistical-learning}, \cite[p.129f]{Goodfellow-et-al-2016}. \newline
Furthermore, due to our choice of an additional sigmoid layer, the output of the network $\hat{y}$ is already a probability distribution over the two possible values $\hat{y} \in \{0,1\}$. Thus, it is only a natural decision to consider the cross-entropy function as our cost function. 

\subsection{Optimizer}
An optimizer, as the name already suggests, tries to optimize the gradient descent approach by regulating the respective step sizes and computing them individually for the parameters.
One of these stochastic gradient descent optimizers is ADAM (derived from Adaptive Momentum Estimator) and was proposed by Kingma and Ba. \cite{kingma2014adam}. ADAM computes the adaptive learning rate for different parameters individually by using estimates of the first and second moments of the gradients\footnote{Moments are quantitative measures of a function that describe the appearance of a distribution \cite[p.267]{Blitzstein2014}}. One of the attractive properties of ADAM is that each time step, the step size is approximately bounded \cite[p.1f]{kingma2014adam} by the initial step size parameter such that a so-called \say{trust region \cite[p.3]{kingma2014adam}} is formed around the current parameter values allowing for an automatic annealing process \cite{kingma2014adam}.
\newline
The necessary parameters for the algorithm, $\alpha$, $\beta_1$, $\beta_2$, $\epsilon$ are initialized with $\alpha=0.001$, $\beta_1=0.9$, $\beta_2=0.999$ and $\epsilon=10^{-8}$ as suggested by \cite{kingma2014adam}.

\section{Results}
Before we come to the results of the experiment, we first have to clarify how results are measured. Especially in medical data, which is often imbalanced, special precaution has to be taken while analyzing and evaluating because slightly different outcomes can have entirely different consequences. Therefore, in the following, it is stated, why the classic paradigm of accuracy as a performance measure fails in our context and which alternative metric is taken. Then, the results of the experiment are presented, based on the introduced metrics. 

\subsection{Evaluation Process}
As the goal of this thesis is to ultimately construct a model that can differentiate BrS positive from BrS negative patients, it is of utmost importance that the classifier can perform well on medical evaluation methods.  
\newline
Whenever the classifier\footnote{In the following, the situation for binary classifiers are described. It is possible to extend the information given to a multi-classifier-case. However, this will not be part of this thesis.} decides whether the current instance belongs to a class or not, we can evaluate its choice by comparing the output $\hat{y}$ with the target output $y$.
\newline
For our purposes, the accuracy\footnote{Accuracy $= \frac{\text{TP}+\text{TN}}{\text{P}+\text{N}}$, where TP stands for the true positives, TN for true negatives, P for the total amounts of positives and N for the total amount of negatives.} gives a first impression of the performance of our model as \say{the ability of the test to distinguish between the relevant states or conditions of the subject (i.e., diagnostic accuracy) is the most basic property of the test as an aid in decision making \cite{Zweig1993ReceiveroperatingC}.} For the tests presented in the following, the classification threshold\footnote{Please note that this classification threshold is picked rather arbitrary. Depending on the classification task it is very reasonable to consider an expert's opinion. E.g., a physician ought to be careful when it comes to making a diagnosis and a value of $0.4$, which is not sufficient for a positive classification in our case, might give a good reason to conduct another screening. This is very well described in \url{https://www.fharrell.com/post/classification/} [date of last access: 05.02.2020].} is set as $0.5$ for measuring the accuracy value: 
\begin{align}
\hat{y}(x)=\begin{cases} 1 & \text{if } \hat{\sigma}(x) > 0.5 \\ 0 &\text{else}\end{cases}
\end{align}
\smallskip

where $\hat{\sigma}(x)$ denotes the final output of the network (after the last sigmoid layer). Nevertheless, accuracy alone does not give a sufficient account of our performance. Since the data sets are highly unbalanced, good accuracy can be achieved by labeling every instance as negative which would mean that our model did not learn anything about the positive examples. This would be a worst-case scenario as our goal is to build a classifier that can detect BrS positive patients.
\newline
Therefore, receiver operating characteristic (ROC) \cite{FAWCETT2006861}, \cite{Zweig1993ReceiveroperatingC}, \cite{Powers2008} is very important. The ROC is a long-established method for \say{visualizing, organizing and selecting classifiers based on their performance \cite[p.1]{FAWCETT2006861}} in the medical field, e.g Zweig et al. \cite{Zweig1993ReceiveroperatingC} or Zou \cite{Zou} (an online library about ROC literature research in medicine) and lately, it is gaining more attention in the domain of machine learning, to illustrate \cite{eFerri2002}, \cite{Provost2001}, \cite{Bradley97}. For the ROC, the true positive rate (sensitivity) is plotted against the false positive rate (1-specificity) and thus gives an account on the performance of our classifier given the ratio of correctly and incorrectly classified positive instances. Using the ROC, different decision thresholds are calculated and used to test the performance of the classifier \cite{FAWCETT2006861}. Each point on the ROC-space stands for a different selected decision threshold and connecting these yields a continuous curve. Depending on where the curve (or the points) is (or are) situated, we can evaluate whether a good classification occurred (upper left area) or a bad classification occurred. E.g., close to the diagonal from bottom left to the upper right, in which case the classifier appears to be making random decisions. 
\newline 
By making use of the true positive rate and false positive rate, and thus ratios, as compared to absolute values, the evaluation of the model is insensitive to an imbalance in the class distribution \cite{FAWCETT2006861}. Hence, the ROC helps us to see whether the accuracy of the model is based on the imbalanced data distributions or its performance.\newline
Calculating the area under the ROC-curve yields the AUC (area under the curve), another metric with whom we are able to compare
the performances of different classifiers by comparing their respective AUCs (as they are scalar values) \cite{Powers2008}.

\subsection{Parameter Search}
Most of the hyperparameters, such as the number of units in the hidden layer or the learning rate, have to be found by a search through the parameter space. To find the best settings, one has to conduct many experiments with different parameter tunings and compare their results. To find suiting parameters, the different models are first trained with the training set and then their performance is tested by using the validation data set.  \newline
As aforesaid, during the process of writing this thesis, time and resources are very limited. Thus, only a small search is performed and some parameter settings are adjusted based on a search through the literature. In total, $322$ experiments are conducted, searching among a possible number of epochs ($5, 10, 15$), the amount of hidden units ($100, 150, 200$), the number of stacked LSTM-layers ($1, 2, 3$) and different learning rates ($10^{-2}, 10^{-3}, 10^{-4}, 10^{-5}$) for each of the different tested features (V1, V2, V1 \& V2). Greff et al. \cite{Greff15} gave the learning rate a high priority together with a suggestion to start high and then divide the learning rate each time by a factor of $10$. Therefore, it is tested with four different settings, each time, dividing by a factor of ten. Next, the amount hidden units follows a suggestion by \cite{LSTMHyperparameters} - after performing a hyperparameter search for different data sets - that gave the number of hidden units a low priority and suggested $100$ units per LSTM-layer\footnote{In their experiment, they used a bi-directional LSTM-model which is not subject to this thesis. However, their suggested amount of units appears to be reasonable and serves as a guideline for the presented experiments.}. According to Reimers and Gurevych \cite{LSTMHyperparameters}, the number of LSTM-layers stacked on top of each other has a medium priority with a suggested amount of $2$. The possible numbers of epochs stem firstly from the restricted amount of computational power and secondly, due to the limited amount of data available for the trials, to avoid overfitting to the training data.  
\newline

\newpage
\begin{figure}[H]
    \begin{subfigure}{\linewidth}
    \hspace*{-0.75cm}
    \includegraphics[scale=0.45]{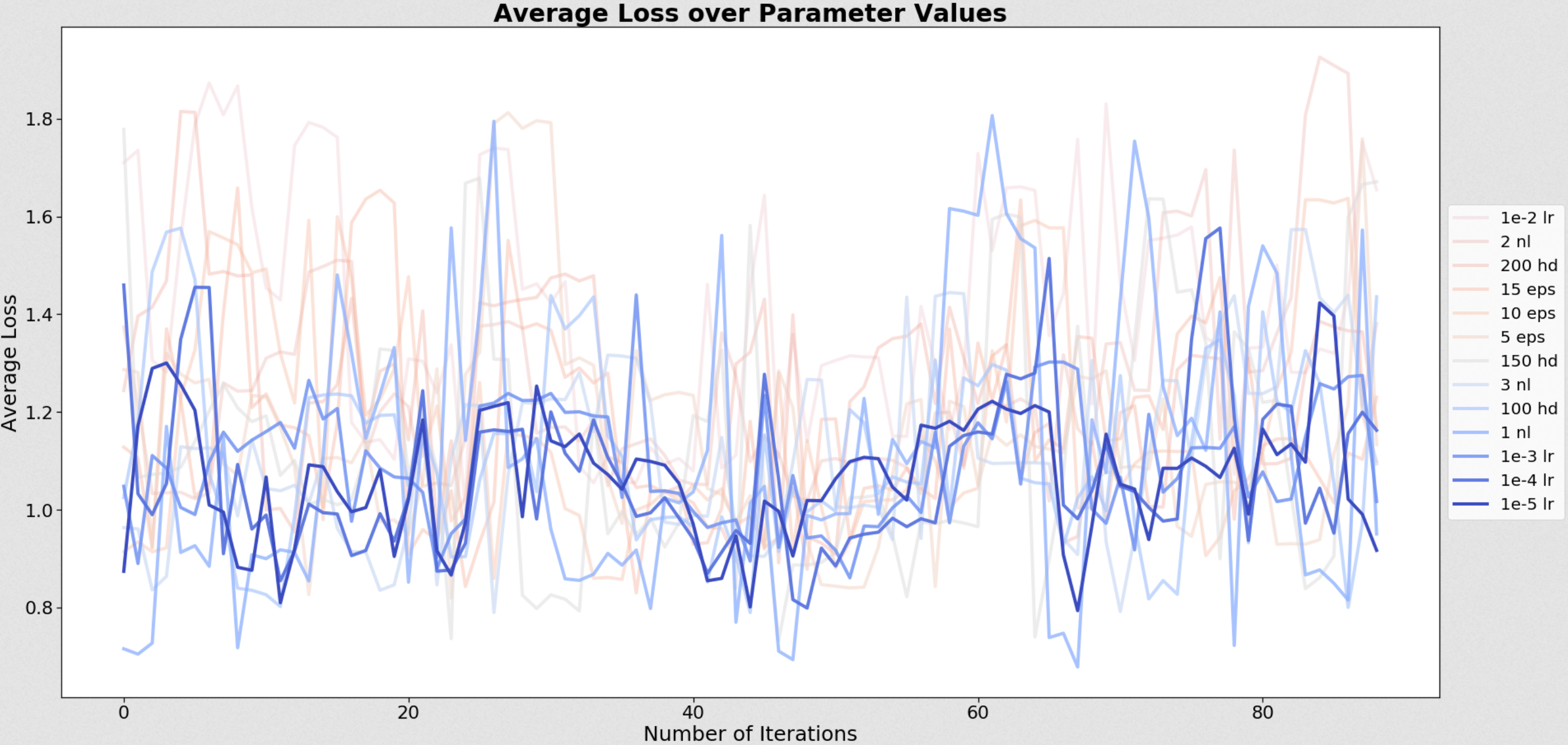}
    \centering
    \caption{}
    \vspace{0.5cm}
    \end{subfigure}\par\medskip
    \begin{subfigure}{\linewidth}
    \includegraphics[scale=0.75]{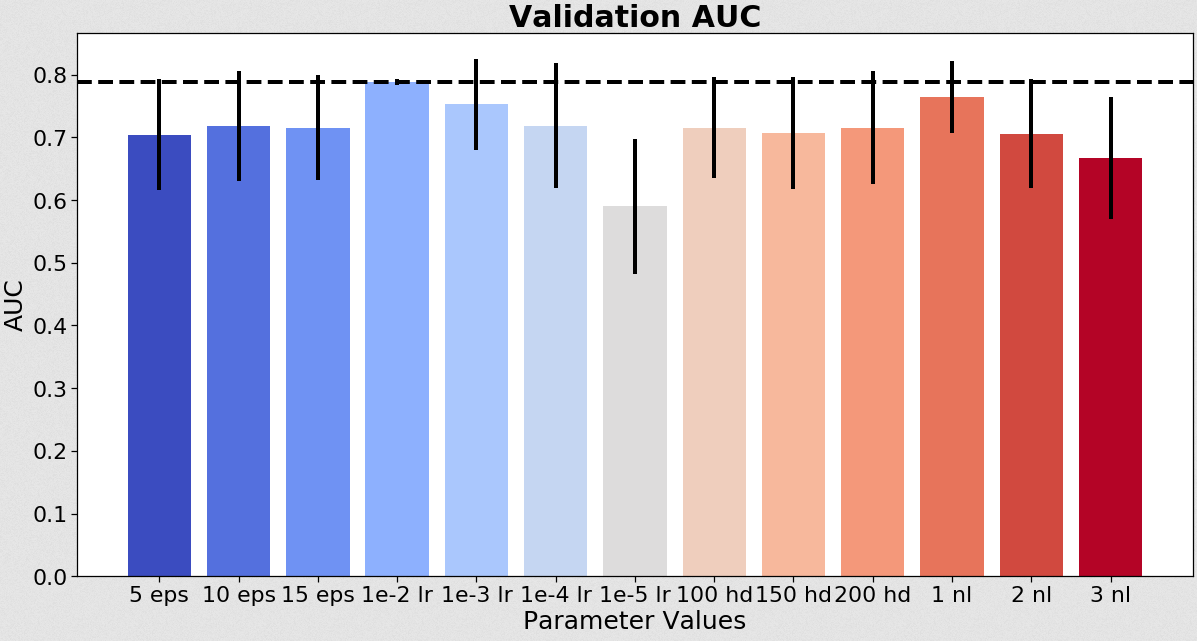}
    \centering
    \caption{}
    \end{subfigure}
    \caption{The parameter search for the model based on the V1 lead, where one value is fixed and then the (a) loss (sorted after blue for lowest and red for highest total loss) and (b) AUC (with confidence intervals) are averaged over all remaining model combinations.}
    \label{fig:AllParams}
    \centering
\end{figure}

The search is conducted as depicted in Figure \ref{fig:AllParams}. After training the models, the average values for the tested metrics (accuracy, AUC, loss) are calculated for every possible parameter value by holding the parameter value fixed and then averaging over all other possible combinations. To illustrate, when investigating the performance for parameter $epochs \in [5, 10 ,15]$, one value is chosen, i.e., $5$, and then for every other possible combination - $learning\_rate \in [10^{-2}, 10^{-3}, 10^{-4}, 10^{-5}]$, $hidden\_dim \in  [100, 150, 200]$ and $num\_layers \in [1, 2, 3]$ - models are constructed and their results are averaged. The reasoning behind this procedure is firstly to speed up the evaluation process and secondly to find the parameter value that seems to have the biggest impact on the performance of the model. Hereby, the AUC of the validation set and the validation loss are the major criteria for choosing a parameter. Only then, accuracy is taken into consideration. 
\begin{figure}[H]
    \centering
    \begin{minipage}{0.7\textwidth}
        \centering
        \includegraphics[width=0.9\textwidth]{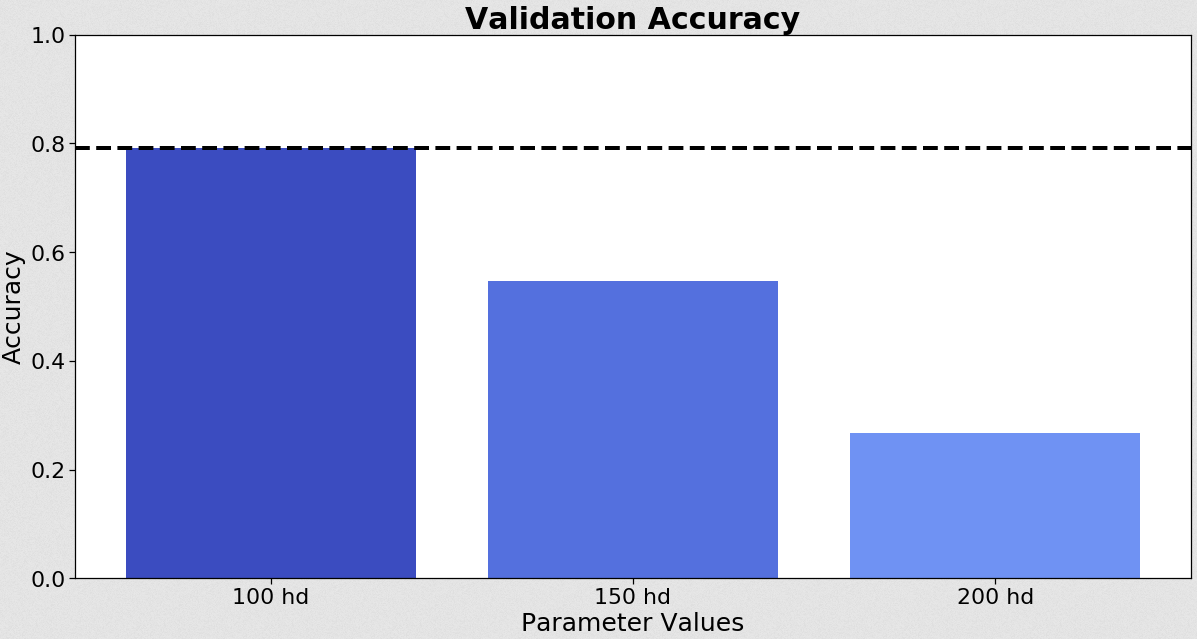} 
        \caption{Validation accuracies for the last parameter setting after fixing the prior three.}
        \label{fig:lastP1}
        \vspace{0.5cm}
    \end{minipage}
    \begin{minipage}{0.7\textwidth}
        \centering
        \includegraphics[width=0.9\textwidth]{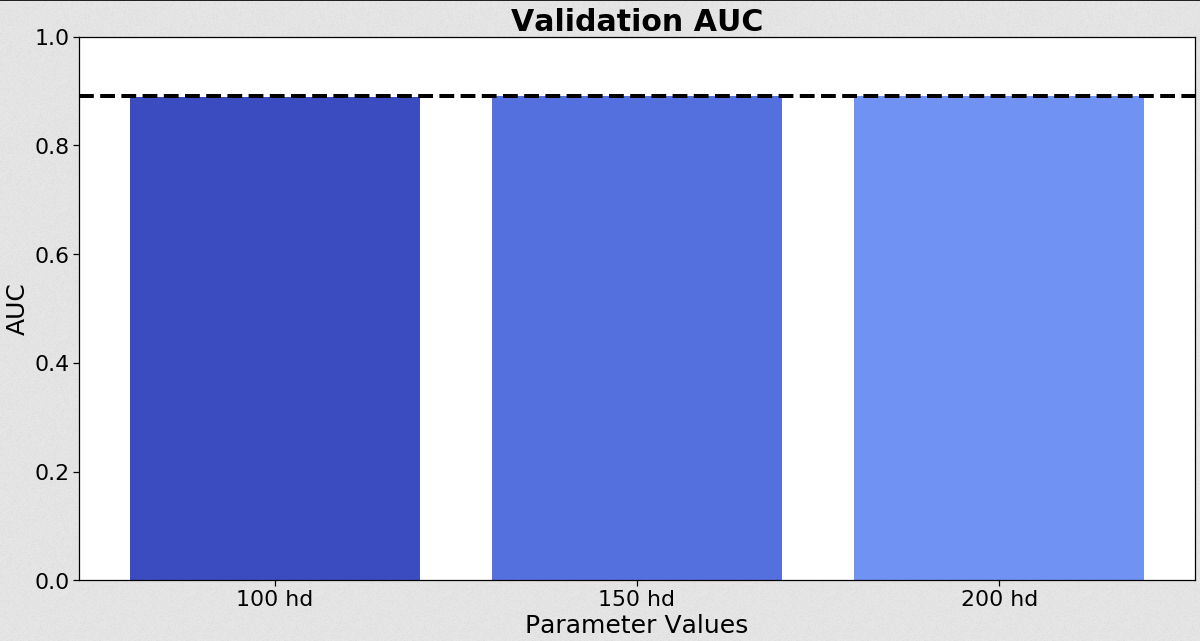} 
        \caption{Validation AUCs for the last parameter setting after fixing the prior three.}
        \label{fig:lastPar2}
    \end{minipage}
\end{figure}

As it can be inferred from Figure \ref{fig:lastP1} and Figure \ref{fig:lastPar2}, this procedure is repeated after deciding for the first parameter value, and then the second, etc. until the last parameter value for the last parameter is found.
\newline
The outcome is three different models, one based on V1-data, one model based on V2-data and the third one based on V1- and V2-data. 
\begin{itemize}
    \item V1: $epochs=15$, $learning\_rate=10^{-3}$, $hidden\_dim=150$,\\ $num\_layers=1$
    \item V2: $epochs=10$, $learning\_rate=10^{-3}$, $hidden\_dim=100$,\\ $num\_layers=1$
    \item both: $epochs=10$, $learning\_rate=10^{-3}$, $hidden\_dim=100$,\\ $num\_layers=1$
\end{itemize}

\subsection{Test Results}
After the decision was made for the fitting parameter values, the chosen models were tested against each other.

\begin{figure}[H]
    \includegraphics[scale=0.75]{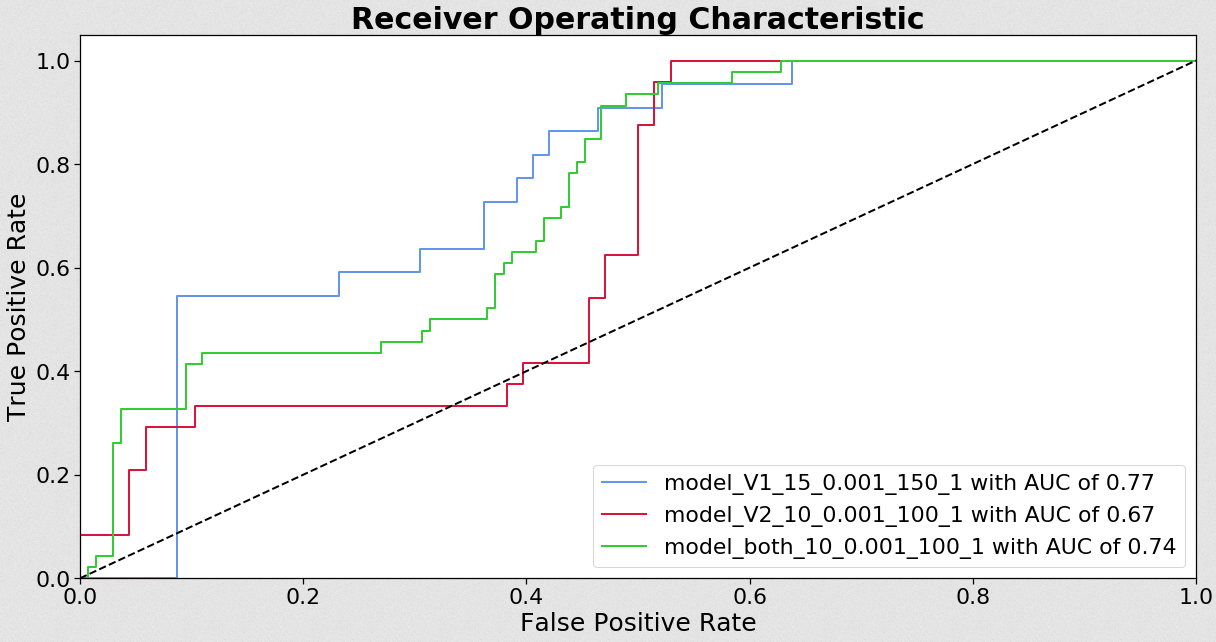}
    \caption{The ROC-curves of the chosen models plotted against each other.}
    \centering
    \label{fig:ROCplots}
\end{figure}

Figure \ref{fig:ROCplots} shows the ROC curves of the models after they encountered and classified the data out of the test set. V1 yielded the highest AUC, closely followed by the combination of V1 \& V2 and lastly, V2. 
\smallskip

\begin{figure}[H]
    \centering
    \begin{minipage}{0.45\textwidth}
        \centering
        \includegraphics[width=0.9\textwidth]{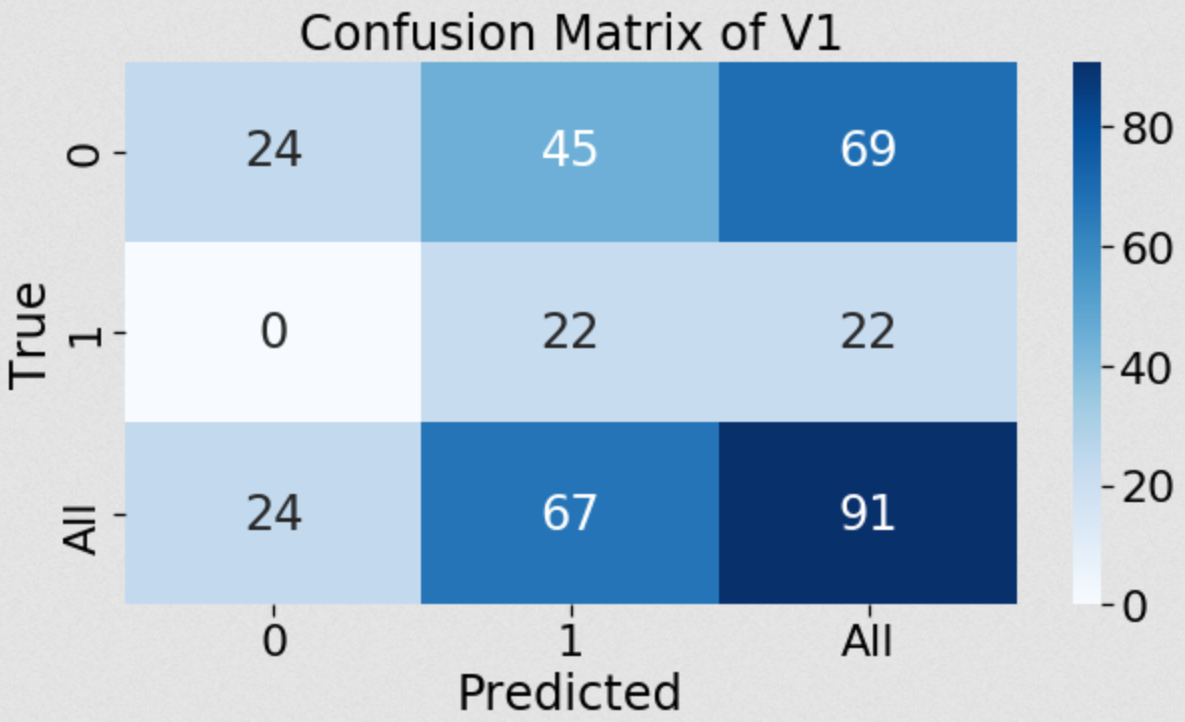} 
        \caption{Confusion matrix of the model that was solely trained on V1 data.}
        \label{fig:cmV1}
    \end{minipage}\hfill
    \begin{minipage}{0.45\textwidth}
        \centering
        \includegraphics[width=0.9\textwidth]{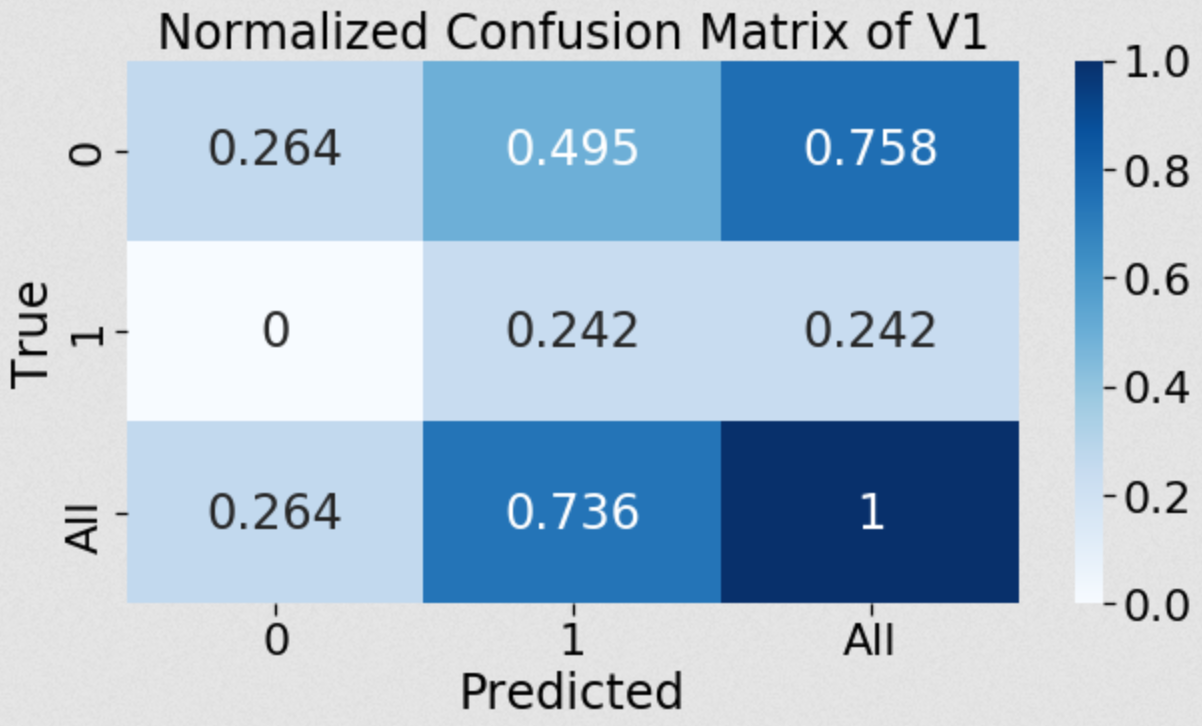} 
        \caption{Normalized confusion matrix of the model that was solely trained on V1 data.}
        \label{fig:NcmV1}
    \end{minipage}
\end{figure}

\begin{figure}[H]
    \centering
    \begin{minipage}{0.45\textwidth}
        \centering
        \includegraphics[width=0.9\textwidth]{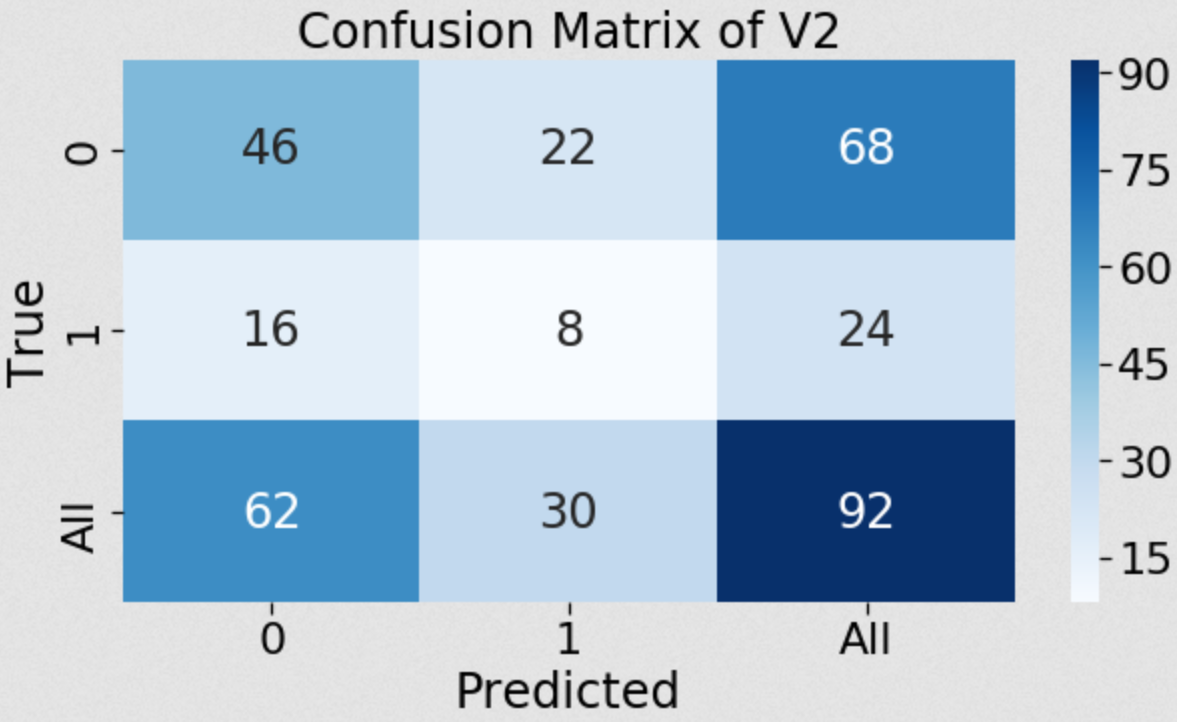} 
        \caption{Confusion matrix of the model that was solely trained on V2 data.}
        \label{fig:cmV2}
    \end{minipage}\hfill
    \begin{minipage}{0.45\textwidth}
        \centering
        \includegraphics[width=0.9\textwidth]{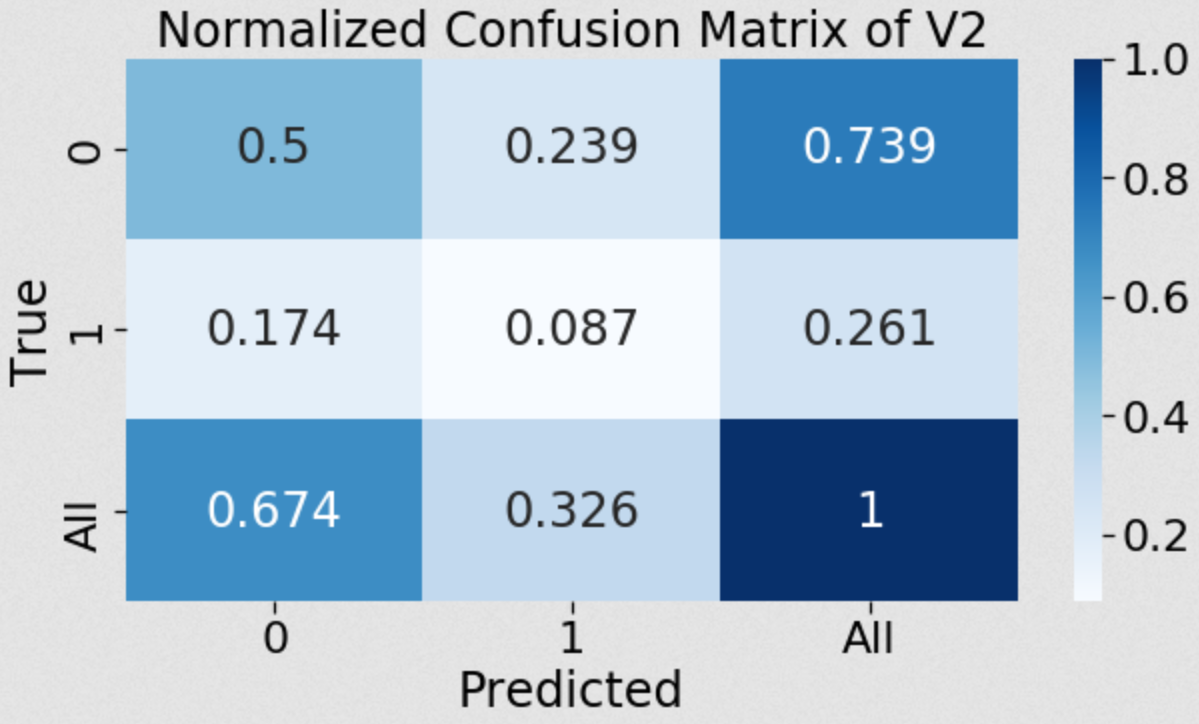} 
        \caption{Normalized confusion matrix of the model that was solely trained on V2 data.}
        \label{fig:NcmV2}
    \end{minipage}
\end{figure}

\begin{figure}[H]
    \centering
    \begin{minipage}{0.45\textwidth}
        \centering
        \includegraphics[width=0.9\textwidth]{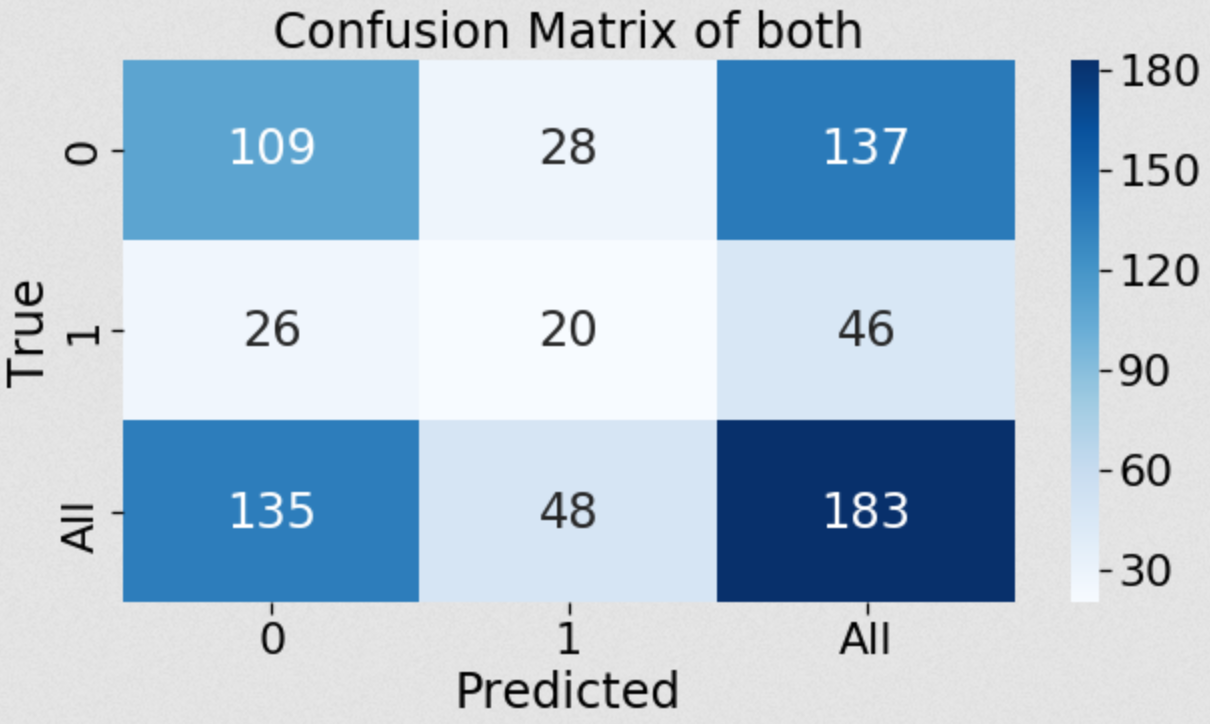} 
        \caption{Confusion matrix of the model that was trained on V1 \& V2 data.}
        \label{fig:cmBoth}
    \end{minipage}\hfill
    \begin{minipage}{0.45\textwidth}
        \centering
        \includegraphics[width=0.9\textwidth]{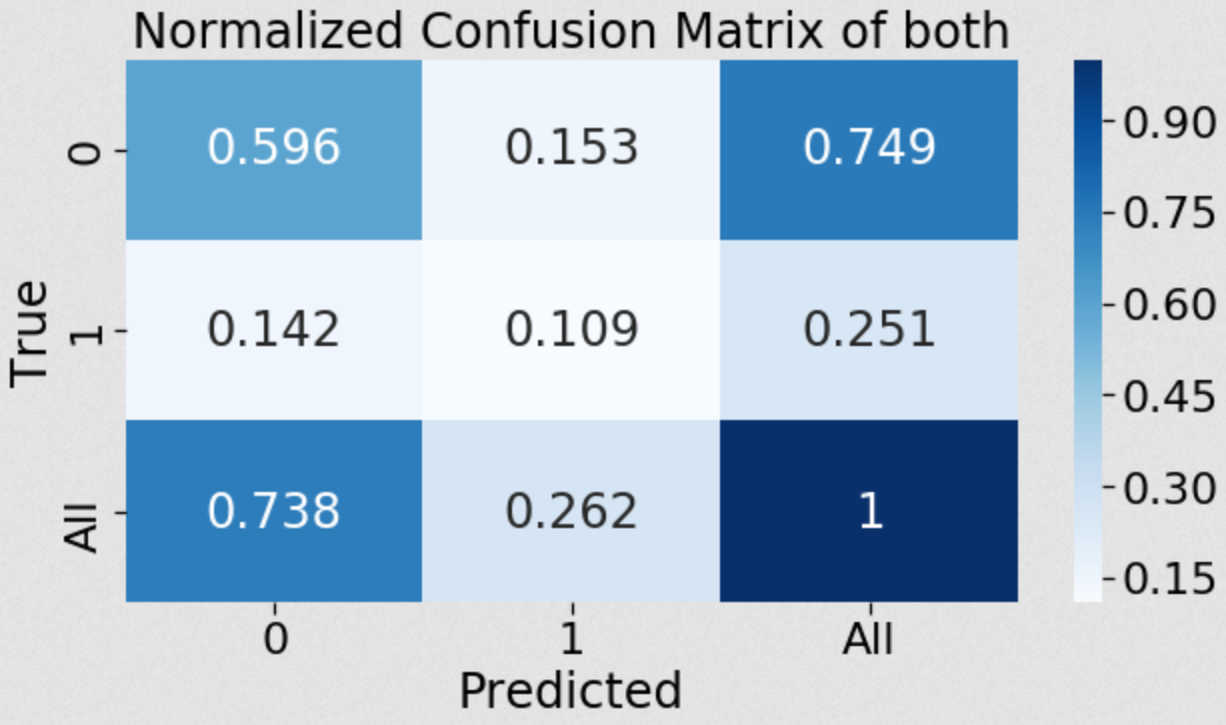} 
        \caption{Normalized confusion matrix of the model that was trained on V1 \& V2 data.}
        \label{fig:NcmBoth}
    \end{minipage}
\end{figure}

The respective confusion matrices\footnote{Bear in mind that the confusion matrices are calculated using the classification threshold that was chosen to be $0.5$ and rather arbitrary. Therefore, they do not give as much as information as the ROC plot does.} and their normalized versions are depicted from Figure \ref{fig:cmV1}-\ref{fig:NcmBoth}. True negatives of the testing procedure are placed in the upper left square, false positives in the row below the true positives. The second column consists of the false negatives on top, true positives in the middle and total of classified positive instances in the bottom. True positives and false negatives give account over the number of positive instances encountered (second row, rightmost square) while false positives and true negatives (first row, rightmost square) yield the number of negative examples. 
\newline 

\begin{table}[h!]
\centering
\begin{tabular}{ c c c c }
 \hline
 Metrics & V1 & V2 & V1 \& V2 \\
 \hline
 \hline
 Val. Total Loss  &  $79.63$    &$73.31$&   $154.83$\\
 \hline
 Val. AUC&   $0.79$  & $0.89$   &$0.83$\\
 \hline
 Val. ACC &$0.47$ & $0.79$ &  $0.81$ \\
 \hline
 Test AUC    &$0.76$ & $0.66$&  $0.74$\\
 \hline
 Test ACC&$0.50$ & $0.58$ & $0.70$\\
 \hline
\end{tabular}
\caption{Results of the selected Models}
\label{tbl:01}
\end{table}

Table \ref{tbl:01} presents the validation and test metrics that were tested on the selected models. Note that the decision threshold was set at $0.5$ and therefore, the accuracy value is not fully reliable. Furthermore, the total validation loss of the V1 \& V2-model is, of course, higher, because it saw twice the amount of ECGs during the validation step. 

\section{Discussion}
Throughout the preceding chapters, instructions and their justifications were given of how one can construct a classifier based on the data that was previously gathered via the digitization process of Part I. This classifier was tested and the results were presented in the previous chapter. We will now interpret these results and then come to a final conclusion after proposing future research.
\newline

By inspecting Figure \ref{fig:ROCplots}, we can tell that the classifiers seem to make more liberal choices, meaning that they risk a higher false positive rate to achieve a higher true positive rate at the same time. In our context of the heart disease, this means that they are quicker to classify BrS-negative ECGs as positive in order to find more BrS-positive ECGs. By analyzing the plot, it appears that they need less evidence in order to make a positive prediction. 
Figure \ref{fig:cmV1} and Figure \ref{fig:NcmV1} illustrate nicely, that a high amount of false positives (almost 50\%) were generated, while yielding no false negatives. Interestingly, the model that is based on the combined data of lead V1 \& lead V2 did make predictions that led to false negatives. Thus, it has misclassified some true BrS-positive ECGs. Even to such an extent that the proportion of true positives captured by the model is lower than the proportion of false negatives. In our medical context, it would mean that the classifier would not be a good advisor for the physician, as it would miss more BrS-positive patients that it could find them. \newline
Table \ref{tbl:01} summarizes the metrics that were tested on all models. It is interesting to see that for V1 the testing accuracy is exactly the chance level, however, the AUC is not. Therefore, the chosen decision threshold of $0.5$ does not work for this model. The other two models are performing slightly better than the model based on V1 in this sense, however, their accuracy's as well as their AUC's are declining when comparing validation and test results. Thus it seems likely that the models did overfit to the training data. In comparison, the V1 model has a relatively marginal decline from validation to test AUC.
\newline
If someone had to make a decision about which model to choose, one would be best advised to take the model based on V1-data because it will be better to have some false positives and run more tests with them, than having false negatives when facing such a severe disease.
\newline

The results of the experiment give the impression that the presented classifiers are not fully developed yet. They will not be a flawless advisor for physicists when they are about to take their final decision upon a diagnosis.
\newline
However, the results are fruitful enough to show that it is possible to prove the concept and that we successfully came closer to our final goal, namely, to construct a classifier that is able to distinguish between BrS positive and BrS negative patients such that it will become an aid to the physicians. It will be now important to consult an expert's opinion to conduct proper validation. Then, future work will hopefully lead to further improvement.  
\newline

In future approaches, the goal will be to achieve a ROC plot, where the curve sticks to the upper left corner right from the start and stays high throughout the threshold testing. Realizing this will result in a classifier that is able to be a valid consultant when making a diagnosis of the BrS. Improvements could be made by considering only data, that was sampled either by the means of Part I, or that was directly sampled from digital ECGs. Standardizing the data gathering process will lead to data that is more closely related such that the model has to make finer distinctions while classifying and hopefully, will learn this during the training process. 
\newline
A prolonged and broader parameter search will results in an abundance of different models not only giving account over the impact of different parameters but also their respective optimal settings. Once the computational infrastructure is given, this will help tremendously in finding an improved classifier.  
\newline
Furthermore, the ECG signals were fed into the classifier as raw time-series data. One could think of various different techniques taken out of the domain of signal processing that exists to shape the data priorly. Future experiments will show whether these techniques - i.e., wavelet transforms that decompose a signal into subparts of itself by passing it again and again through low and high pass filters each time dividing the signal further, giving account over which frequency is present at which time \cite{PolikarWaveLet} - are useful data preprocessing tools to achieve better results. 
\newline
Moreover, after constructing a good model, it will be important to find out what the model actually has learned. This will give further insight into the decision making and may spark up the investigation for new methods of diagnostics. Remarkably, van der Westhuizen, Jos \& Lasenby \cite{van2017visualizing} have recently shown that the hidden neurons in LSTM networks seem to \say{extract representations from the frequency domain similar to wavelet transformations \cite[p.8]{van2017visualizing}.} Thus, preprocessing the data might not lead to an improvement after all but this will be another topic of research.
\newline
Another promising future research will be to investigate the strategies of Oh et al. \cite{OH2018278} who first made use of a convolutional neural network to extract features from the ECG signal and then subsequently an LSTM network for classification. Again, it will be very interesting to study the decision making of the combined network to extract further information for physicists. 
\newline

Overall, the presented experiments were successful by providing a blueprint for the construction of a classifier regarding the BrS. A proof of concept has been stated and enough fuel for future investigations was provided. 

\section{Conclusion}
This second part of the thesis dealt with the possibility of constructing a classifier that was based on the data gathered during Part I of the dissertation. The aforementioned classifier is built by using the LSTM architecture and its parameters were found by a search through the space of possible parameter values. The final results were presented and interpreted, giving an extensive account of the performance of the presented models. Finally, future work was proposed that will give insight into how the presented model could be further augmented. \newline
Our goal to state a proof of concept has been successfully reached.

\chapter{Closing Remarks}
In this thesis, we have demonstrated that it is possible to create a classifier that can diagnose the BrS based on ECG signals and thus, providing help in the need for clear diagnostics. \newline

The proposed pipeline to digitize the ECG signals distinguished between three different types of ECG images that were encountered throughout the project. First, the background is removed as well as other non-relevant pieces of information. Then, the signal was extracted into a 1D vector, mapped into time-voltage coordinates and finally upsampled. The final reconstruction of the ECG signal is very close to the original signal on the ECG strip, although in some cases, the extracted signal was harmed due to residuals or a too aggressive removal process. \newline
The data gathered by the pipeline functioned as the input to the presented machine learning model. This model was constructed based on the LSTM, a special type of neural network. Architectural choices were displayed and the necessary processing steps described. Furthermore, a parameter search was conducted to find the models that yielded the best diagnostic performance. In the end, a classifier was found that proofed the concept of diagnosing BrS via machine learning tools solely based on digital data.
\newline

With the here presented work, it is hoped to achieve further progress in the research regarding the BrS and to lay out a first blueprint on how computational methods will play a big role in the diagnostics of the disease.

\newpage
\pagenumbering{roman}
\setcounter{page}{9}
\bibliography{main.bib}
\bibliographystyle{ieeetr}

\end{document}